\documentclass[aps,pre,letterpaper,twocolumn,final,superscriptaddress,floatfix]{revtex4-2}
\usepackage{lineno}

\usepackage{graphicx,psfrag,amsmath,amssymb,amsfonts,bbm,latexsym,color,dcolumn,bm,mathbbol}
\usepackage[framemethod=tikz]{mdframed}
\usepackage[colorlinks=true,allcolors=blue]{hyperref}

\usepackage{soul} 
\usepackage{cancel}
\usepackage{comment}

\usepackage{array}
\usepackage{booktabs}
\usepackage{footnote}
\usepackage{threeparttable}

\usepackage{natbib}
\usepackage{amsmath,amssymb}             
\usepackage{color}

\newcommand{\cellJ}{\text{cell}\kern0.08em J}

\usepackage{enumerate}
\usepackage[normalem]{ulem}

\definecolor{jfrcolor}{rgb}{0.8, 0, 0.2}

\definecolor{szlcolor}{rgb}{0.75, 0, 0.75}

\definecolor{bluecolor}{rgb}{0, 0, 1}

\begin{document}


\title{Viscous vertex model for active epithelial tissues}

\author{Shao-Zhen Lin}
\email{linshaozhen@mail.sysu.edu.cn}
\affiliation{Guangdong Provincial Key Laboratory of Magnetoelectric Physics and Devices, School of Physics, Sun Yat-sen University, Guangzhou 510275, China.}
\affiliation{Interdisciplinary Research Center for Physical Mechanics in Complex Systems and Its Engineering Applications, School of Physics, Sun Yat-sen University, Guangzhou 510275, China.}
\author{Sham Tlili}
\affiliation{Aix Marseille Univ, IBDM (UMR 7288), Turing Centre for Living Systems, Marseille, France}
\author{Jean-Fran\c{c}ois Rupprecht}
\email{jean-francois.rupprecht@univ-amu.fr}
\affiliation{Aix Marseille Univ, Université de Toulon, CPT (UMR 7332), Turing Centre for Living Systems, Marseille, France}
\affiliation{Aix Marseille Univ, CNRS, LAI (UMR 7333), Turing Centre for Living Systems, Marseille, France}

\begin{abstract}
We present a rotationally invariant viscous vertex model that accounts for both cortical and bulk dissipation of cells. The vanishing substrate-friction limit is enforced via Lagrange multipliers, which also provides a framework for implementing various boundary conditions, such as fixed boundaries and prescribed tractions. Building on this formulation, we introduce a slab-shear rheology protocol to extract an effective, coarse-grained tissue shear viscosity. Under polar or nematic activity, viscosity regulates the formation of elongated, spatially correlated cell-shape textures and stabilizes well-defined topological defects. Because the model remains well-posed at zero substrate friction, it is naturally suited to describing free-floating epithelia and organoids.
\end{abstract}

\date{\today}

\maketitle

\section{Introduction}

Confluent epithelial monolayers and other dense cell assemblies behave as active viscoelastic materials \cite{Forgacs1998,Tlili2020,Fu2024,Khalilgharibi2019,Arora2025,Karnat2025,Nishizawa2023}. 
On short time scales ($\lesssim 1 \ \rm s$), cell junctions and cortical networks respond elastically to imposed deformations and forces, whereas over minutes, junctional crosslinkers turn over and actomyosin structures remodel, enabling sustained cortical flows under applied stresses \cite{Forgacs1998,Feng2025,Cheng2025,Clement2017,Sun2026}. 
This rheological behavior underpins a wide range of morphogenetic processes and collective cell migration phenomena \cite{Fu2024,Clement2017}. 
It is often described at a continuum scale within active polar/nematic gel frameworks, where effective viscosities and active stresses are introduced phenomenologically \cite{Prost2015,Shaebani2020,Shankar2022}. 

\begin{figure}[t!]
\includegraphics[width=8.6cm]{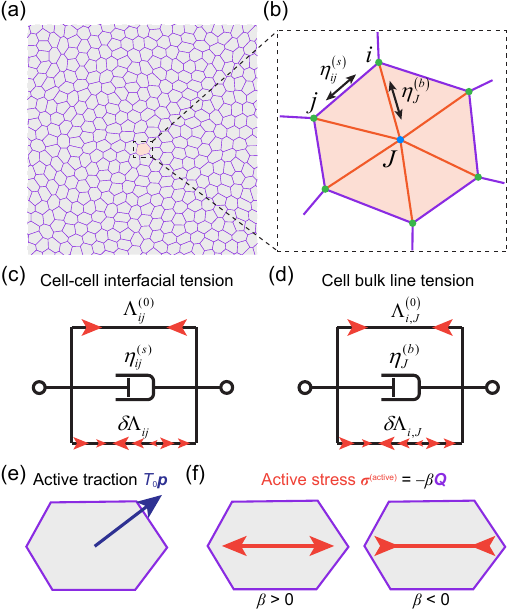}
\caption{\label{fig:Model} 
Schematic of the active, viscous vertex model. 
(a) In vertex models, a cell sheet is mimicked by a tiling of polygons. 
(b) Schematic of the two kinds of cellular viscosities considered in our model, including the cell--cell interfacial viscosity ($\eta_{ij}^{(s)}$) and the cell bulk viscosity ($\eta_J^{(b)}$). 
(c) Schematic of the cell--cell interfacial tension of the interface $ij$, including three contributions: (1) a constant tension $\Lambda_{ij}^{(0)}$; (2) a viscous tension $\eta_{ij}^{(s)}\dot{\ell}_{ij}$; (3) tension fluctuations $\delta \Lambda_{ij}$. 
(d) Schematic of the cell bulk line tension of the segment $iJ$, including three contributions: (1) a constant tension $\Lambda_{iJ}^{(0)}$; (2) a viscous tension $\eta_{J}^{(b)}\dot{\ell}_{i,J}$; (3) tension fluctuations $\delta \Lambda_{i,J}$. 
(e) Schematic of the polar active traction force $\bm{F}_J^{(\rm active)} = T_0 \bm{p}_J$ with $T_0$ being the active traction magnitude and $\bm{p}_J$ the cell polarization vector. 
(f) Schematic of the apolar active stress $\bm{\sigma}^{\rm (active)}_J = -\beta \bm{Q}_J$ with $\beta$ being the cellular activity parameter and $\bm{Q}_J$ the cell shape anisotropy tensor. 
}
\end{figure}

Cell-based models, particularly vertex models, provide a complementary and more microscopic description of tissue mechanics. In standard vertex models, cells are represented as polygons (Fig. \ref{fig:Model}(a)) whose vertices move in an overdamped manner in response to forces derived from area and perimeter elastic energies, together with prescribed cell--cell interfacial tensions \cite{Fletcher2014,Bi2015,Lin2017,Lin2018,Lin2023,Chen2022}. Dissipation is commonly modeled as a local friction between vertices and a static mechanical support, such as the extracellular matrix or a rigid substrate \cite{Alt2017}. In this widely used formulation, the junctional cortex is effectively treated as an elastic element, while viscous contributions to junctional tension are neglected.

Yet internal viscous forces dominate over external friction in several systems, e.g., within free-floating embryonic tissues and free-standing monolayers \cite{Harris2012,Armon2018,Duque2024}. 
In the \textit{Drosophila} embryo, estimates of tissue flows and force balances indicate that specific friction against the surrounding medium is often small compared to bulk and cortical viscosities \cite{DAngelo2019,Doubrovinski2017,Rupprecht2018}. 
The viscosity of the cell cortex (resp. the cytoplasm) stems from multiple microscopic, dissipative processes beneath the membrane (resp. in the crowded intracellular fluid), e.g., the dynamic organization and deformation of F-actin filaments \cite{Simon2019,Najafi2023}. 
At the cellular level, the actomyosin cortex is significantly more viscous than the cytoplasm: the cortical viscosity is on the order of $\eta_{\mathrm{cort}} \simeq 10^5 \  \mathrm{Pa}\cdot\mathrm{s}$ \cite{Etienne2015b}, whereas the cytoplasmic viscosity is only $\eta_{\mathrm{cyt}} \simeq 2 \ \mathrm{Pa}\cdot\mathrm{s}$ \cite{Wirtz2009,Bambardekar2015}. 
Although the cortical layer is thin (typically $\ell_{\mathrm{cort}} \simeq 0.2 \ \mu \mathrm{m}$) compared with the cytoplasm (typically $\ell_{\mathrm{cyt}} \simeq 5 \ \mu \mathrm{m}$), cortical viscous dissipation is about $10^3$ times larger than cytoplasmic viscous dissipation: $(\eta_{\mathrm{cort}} \ell_{\mathrm{cort}}) / (\eta_{\mathrm{cyt}} \ell_{\mathrm{cyt}}) \simeq 10^3$ \cite{Turlier2014}. 
Thus, despite its small thickness, the cortex can provide the dominant contribution to dissipation \cite{Turlier2014,Rupprecht2018,Lenne2021,Sun2026}.

A series of pioneering cell-based studies introduced the $\gamma-\mu$ formalism that includes both a cell bulk viscosity ($\mu$) and fixed junctional tensions ($\gamma$) \cite{Yang2009,Brodland2009}, henceforth accounting for the difference in rheologies at the cell bulk and cortex. 
A recent set of articles then accounted for the possibility of a viscous cortical response, either to describe cell division and apoptosis in suspended monolayers \cite{Okuda2015}, spatial ordering in the \textit{Xenopus} embryo \cite{NestorBergmann2018}, viscous intercellular adhesion in epithelial tissues \cite{Nguyen2026}, or the emergence of spontaneous shear flows in monolayers \cite{Sonam2023, Fu2024, Rozman2025}.

Despite these advances, we still lack a systematic formulation of a viscous vertex model that (i) treats both junctional and bulk viscosities of cells on the same footing, (ii) remains rotationally invariant and independent of arbitrary reference frames, (iii) is well-posed in the limit of vanishing substrate friction, and (iv) yields a clear coarse-grained tissue viscosity.

In this article, we develop a viscous extension of the vertex model that explicitly accounts for both cell--cell interfacial viscosity (i.e., junctional viscosity along cell edges) and cell bulk viscosity (i.e., viscosity between vertices and cell centers), which we refer to as the \emph{viscous vertex model}. Junctional viscosity is introduced through velocity differences projected along edges, and bulk viscosity through relative motion between vertices and cell geometric centers, resulting in a rotationally invariant, frame-independent formulation. The resulting force balance can be written in the following matrix form, 
\begin{equation}
\bm{C} \cdot \bm{v} = \bm{F}^{(\rm t)} , \label{eq:ForceBalance_MatrixForm}
\end{equation}
where $\bm{C}$ is a friction--viscosity coefficient matrix combining substrate friction, junctional viscosity, and bulk viscosity; $\bm{v}$ collects all vertex velocities; $\bm{F}^{(\rm t)}$ gathers all non-dissipative forces (e.g., elastic forces, active forces, stochastic fluctuations, etc). 

Our framework tackles situations in which the friction between cells and their environment is negligible, such as free-floating tissues \cite{Harris2012,Duque2024} and organoids \cite{Gsell2025,Li2024}. As pointed out in Refs. \cite{Staple2012Understanding} and \cite{Alt2017}, in this limit, the coefficient matrix $\bm{C}$ in Eq. \eqref{eq:ForceBalance_MatrixForm} becomes singular because global translations and rotations are unconstrained. This difficulty is resolved by (i) accounting for finite cell bulk viscosity and (ii) augmenting the system into a saddle-point matrix $\bm{C}_{\text{ext}}$, by appending global or boundary kinematic constraints via a Lagrange multiplier formalism. These two operations regularize the linear system and enable simulations at strictly zero friction. The same approach is used to implement prescribed boundary conditions, including dragging a single cell or a cell cluster, fixed or freely sliding boundaries in a confinement geometry (e.g., slab, disk, ring geometries), and external pulling protocols.

Within this framework, we introduce a slab-shear protocol to measure a coarse-grained tissue viscosity directly from vertex velocities and line tensions. In the absence of activity, we extract a short-time tissue viscosity from the instantaneous linear relationship between viscous shear stress and imposed shear strain rate. We derive an analytical expression for the short-time tissue viscosity in the case of a tissue formed of regularly packed hexagonal cells, and find that it scales linearly with the junctional and bulk viscosities even at relatively large values. Numerical simulations confirm that this relation holds in disordered cell packings. Allowing for cell rearrangements under sustained shear, we then extract a long-time tissue viscosity, which we also find to scale linearly with the junctional and bulk viscosities.

Finally, the consequences of cellular viscosity are explored in two typical kinds of active vertex models: (1) a polar active vertex model, which accounts for polar active traction forces at the leading edge of cells; (2) a nematic active vertex model, which implements cell-shape-dependent active stresses. Both kinds of cellular activities drive flows and the dynamics of topological defects. Increasing junctional and bulk viscosities leads to more elongated and spatially correlated cell shapes, modifies the density and organization of $\pm 1/2$ topological defects, and promotes coherent flow structures in both periodic domains and confined geometries. Because the formulation remains well-posed as the substrate friction vanishes, it is directly applicable to free-floating tissues and organoids, thereby providing a bridge between cell-based models and continuum active nematic descriptions of living tissues.

\section{A rotationally invariant formulation of dissipation}

\subsection{Implementation of cellular viscosities}
\label{sec:cell_viscosity}

\subsubsection{Cell--cell interfacial viscosity}

In addition to the standard constant interfacial tension, here we incorporate the viscosity of intercellular junctions, see Fig. \ref{fig:Model}(b). 
We define the viscous intercellular line tension $\Lambda_{ij}^{\rm (viscous)}$ based on a dissipation rate at cell--cell interfaces, denoted $S_{\rm interface}$. 
Specifically, we correlate the dissipation rate $S_{\rm interface}$ with the elongation rate (denoted $\dot{\ell}_{ij}$) of cell--cell interfaces. Here, to the lowest order, we express $S_{\rm interface}$ as 
\begin{equation}
S_{\rm interface} = \frac{1}{2}\sum_{\langle i,j \rangle} {\eta_{ij}^{(s)}\dot{\ell}_{ij}^2} , \label{eq:Dissipation_Interface_1}
\end{equation}
where $\eta _{ij}^{\left( s \right)}$ is the viscous coefficient of the cell--cell interface $ij$ (different from the traditional fluid viscosity); ${{\dot{\ell}}_{ij}} = {\text{d}{{\ell}_{ij}}}/{\text{d}t}$ is the elongation rate of cell--cell interface $ij$ with ${\ell_{ij}}=\left| {{\bm{r}}_{i}}-{{\bm{r}}_{j}} \right|$ being the corresponding interface length; the sum $\sum_{\langle i,j \rangle}$ is taken over all cell--cell interfaces $ij$ that connect vertex $i$ and $j$. 
Since ${{\dot{\ell}}_{ij}} = {\text{d}{{\ell}_{ij}}}/{\text{d}t} = {{\bm{t}}_{i,j}}\cdot \left( {{\bm{v}}_{j}}-{{\bm{v}}_{i}} \right)$ with $\bm{v}_i = \mathrm{d}\bm{r}_i / \mathrm{d}t$ being the velocity of vertex $i$ and ${{\bm{t}}_{i,j}} = ({{{\bm{r}}_{j}}-{{\bm{r}}_{i}}})/{{{\ell}_{ij}}}$ a unit vector along the cell--cell interface $ij$, the dissipation rate at cell--cell interfaces, Eq. \eqref{eq:Dissipation_Interface_1}, can be re-expressed as 
\begin{equation}
S_{\rm interface} = \frac{1}{2}\sum_{\langle i,j \rangle} {\eta_{ij}^{(s)}\left[ {{\bm{t}}_{i,j}}\cdot \left( {{\bm{v}}_{j}}-{{\bm{v}}_{i}} \right) \right]^2} . \label{eq:Dissipation_Interface}
\end{equation}

The definition of the dissipation rate at cell--cell interfaces in Eq. \eqref{eq:Dissipation_Interface_1} is equivalent to assuming a linear relation between the cell--cell interfacial viscous tension ${{\Lambda }_{ij}^{(\rm viscous)}}$ and the cell--cell interface elongation rate $\dot{\ell}_{ij}$, that is, 
\begin{equation}
{{\Lambda }_{ij}^{(\rm viscous)}} = \eta _{ij}^{\left( s \right)}{{{\dot{\ell}}}_{ij}} = \eta _{ij}^{\left( s \right)}{{\bm{t}}_{i,j}}\cdot \left( {{\bm{v}}_{j}}-{{\bm{v}}_{i}} \right) . \label{eq:ViscousTension_Interface}
\end{equation}

Note that there are different ways to define the cell--cell interfacial viscous tension $\Lambda^{(\rm viscous)}_{ij}$. 
An alternative, rotationally invariant definition could be $\Lambda^{(\rm viscous)}_{ij} = \eta^{(s)}_{ij}\dot{\varepsilon}_{ij}$, which relies on the cell--cell interface elongation strain rate $\dot{\varepsilon}_{ij} = \dot{\ell}_{ij} / \ell_{ij}$ with $\varepsilon_{ij} = \ln (\ell_{ij} / \ell_{ij,0})$ being the logarithmic strain of the cell--cell interface $ij$. 
In Sec. \ref{sec:alternative_viscous_tension}, we demonstrate that such a definition tends to suppress T1 topological transitions. 
Thus, we focus on the viscous tension definition in Eq. \eqref{eq:ViscousTension_Interface} hereafter. 

Further considering the constant part and fluctuations, the total tension of the cell--cell interface $ij$ can be expressed as (Fig. \ref{fig:Model}(c)), 
\begin{equation}
{{\Lambda }_{ij}} = \eta _{ij}^{\left( s \right)}{{\bm{t}}_{i,j}}\cdot \left( {{\bm{v}}_{j}}-{{\bm{v}}_{i}} \right) + \Lambda _{ij}^{\left( 0 \right)} + \frac{1}{\sqrt{{{\ell}_{ij}}}}{{\zeta }_{ij}} , \label{eq:InterfacialTension}
\end{equation}
where $\Lambda _{ij}^{\left( 0 \right)}$ is a constant tension along the cell--cell interface $ij$; ${{\zeta }_{ij}}$ are stochastic tension fluctuations.

\subsubsection{Cell bulk viscosity}

In addition to the cell--cell interfacial viscosity, we also consider the bulk viscosity of cells, as shown in Fig. \ref{fig:Model}(b). 
Similarly, we define the cell bulk viscosity based on the dissipation rate at the cell bulk, denoted $S_{\rm bulk}$.
In analogy with Eq. \eqref{eq:Dissipation_Interface_1}, we express $S_{\rm bulk}$ as 
\begin{equation}
S_{\rm bulk} = \frac{1}{2}\sum_{J = 1}^{N_c}\sum_{i \in \cellJ} \eta_J^{(b)} \dot{\ell}_{i,J}^2 , \label{eq:Dissipation_Bulk_1}
\end{equation}
where $\eta _{J}^{\left( b \right)}$ is the bulk viscous coefficient of cell $J$; $\dot{\ell}_{i,J} = \mathrm{d}\ell_{i,J}/\mathrm{d}t$ is the elongation rate of the vertex--cell link between vertex $i$ and cell $J$ with ${{\ell}_{i,J}} = \left| {{\bm{r}}_{i}}-{{\bm{r}}_{J}} \right|$ and $\bm{r}_J$ the geometric center of cell $J$; the summation $\sum_{J = 1}^{N_c}$ is taken over all cells indexed by $J = 1, 2, \cdots, N_c$ with $N_c$ being the total number of cells; the summation $\sum_{i \in \cellJ}$ is made over all vertices belonging to cell $J$. 
The geometric center of a cell is defined as ${{\bm{r}}_{J}} = \sum_{k\in \cellJ}{{{\bm{r}}_{k}}} / n_J$ with $n_J$ being the number of vertices (edges) of cell $J$.  
Since $\dot{\ell}_{i,J} = {{\bm{t}}_{i,J}}\cdot ( {{\bm{v}}_{J}}-{{\bm{v}}_{i}} )$ with ${{\bm{t}}_{i,J}} = ({{{\bm{r}}_{J}}-{{\bm{r}}_{i}}})/{{{\ell}_{i,J}}}$ and ${{\bm{v}}_{J}} = {\text{d}{{\bm{r}}_{J}}}/{\text{d}t}$, Eq. \eqref{eq:Dissipation_Bulk_1} can be re-expressed as 
\begin{equation}
S_{\rm bulk} = \frac{1}{2}\sum_{J = 1}^{N_c}\sum_{i \in \cellJ} \eta_J^{(b)} [ \bm{t}_{i,J} \cdot (\bm{v}_J - \bm{v}_i) ]^2 . \label{eq:Dissipation_Bulk}
\end{equation}

The dissipation rate at the cell bulk, Eqs. \eqref{eq:Dissipation_Bulk_1} and \eqref{eq:Dissipation_Bulk}, is equivalent to assuming a virtual link from each vertex $i$ to the geometric center of its neighboring cell and assuming a linear relation between the viscous cell bulk line tension and the vertex-cell center expansion rate, that is, 
\begin{equation}
{{\Lambda }_{i,J}^{(\rm viscous)}} = \eta_J^{(b)} \dot{\ell}_{i,J} = \eta _{J}^{\left( b \right)}{{\bm{t}}_{i,J}}\cdot \left( {{\bm{v}}_{J}}-{{\bm{v}}_{i}} \right) , \label{eq:ViscousTension_Bulk}
\end{equation}
where ${{\Lambda }_{i,J}^{\rm (viscous)}}$ is the viscous line tension between vertex $i$ and its neighboring cell $J$. 

Similarly, considering the constant part and fluctuations, the total tension between cell vertices and the cell center can be expressed as (Fig. \ref{fig:Model}(d)), 
\begin{equation}
{{\Lambda }_{i,J}} = \eta _{J}^{\left( b \right)}{{\bm{t}}_{i,J}}\cdot \left( {{\bm{v}}_{J}}-{{\bm{v}}_{i}} \right) + \Lambda _{i,J}^{\left( 0 \right)} + \frac{1}{\sqrt{{{\ell}_{i,J}}}}{{\zeta }_{i,J}} , \label{eq:BulkTension}
\end{equation}
where $\Lambda_{i,J}^{\left( 0 \right)}$ is a constant vertex-cell line tension and ${{\zeta }_{i,J}}$ are stochastic fluctuations.

\subsection{Motion equation}

\subsubsection{Force balance}

The dynamics of a cell sheet can be characterized by the motion of vertices, whose positions are denoted as $\bm{r}_i$ with $i = 1, 2, \cdots, N_v$ being the index of vertices and $N_v$ the total number of vertices. 
In general, the force balance at vertex $i$ can be decomposed into four generic terms: 
\begin{align}
 \bm{F}_{i}^{\left( \text{elastic} \right)} + \bm{F}_{i}^{\left( \text{active} \right)} + \bm{F}_{i}^{\left( \text{dissipation} \right)} + \bm{F}_{i}^{\left( \text{noise} \right)} = \bm{0} , \label{eq_ForceBalance}
\end{align}
each corresponding to elastic forces associated with cell-shape regulation, to active forces generating motion and deformation of cells, to dissipation resisting motion and cell shape variation, and to stochastic fluctuations, in turn. 
We detail the specific expression of each term within the next paragraphs. 

\textit{Elastic forces.} -- 
We decompose the elastic cell shape relaxation force into three contributions: 
\begin{align}
 \bm{F}_{i}^{\left( \text{elastic} \right)} = \bm{F}_{i}^{\left( \text{area} \right)} + \bm{F}_{i}^{\left( \text{perimeter} \right)} + \bm{F}_{i}^{\left( \text{tension} \right)} , 
\end{align}
which can be obtained by taking the partial derivatives ($- \partial / \partial \bm{r}_i$) of the corresponding mechanical energy of the cell sheet \cite{Fletcher2014, Bi2015, Lin2017p, Lin2018, Lin2023, Chen2022, Lin2017}.
Specifically, the mechanical energy of cell area elasticity can be expressed as $E_{\rm area} = \sum_{J = 1}^{N_c} K_A (A_J - A_0)^2 / 2$, where $K_A$ is the area elastic modulus, $A_0$ is a preferred cell area, and $A_J$ is the area of the $J$-th cell. 
The mechanical energy of cell perimeter elasticity can be expressed as $E_{\rm perimeter} = \sum_{J=1}^{N_c} K_P (P_J - P_0)^2 / 2$, where $K_P$ is the perimeter elastic modulus, $P_0$ is a preferred cell perimeter, and $P_J$ is the perimeter of the $J$-th cell. 
The mechanical energy of tension can be expressed as $E_{\rm tension} = \sum_{\langle i,j \rangle} \Lambda_{ij}^{(0)} \ell_{ij} + \sum_{J = 1}^{N_c} { \sum_{i \in \cellJ} \Lambda_{iJ}^{(0)} \ell_{i,J} }$.

\textit{Active forces.} -- 
The active force $\bm{F}_i^{\rm (active)}$ accounts for forces related to the activity of cells, which typically includes two kinds (Fig. \ref{fig:Model}(e,f)), i.e., the polar active traction force mimicking the activity of cell protrusions at the leading edge of cells \cite{Lin2018,Li2014,Bi2016,Barton2017,Chen2023,Wang2024} and the apolar active cellular stress mimicking the contractility/extensivity of the intracellular cytoskeleton \cite{Lin2023,Sonam2023,Rozman2024,Rozman2025,Yu2025,Yu2024,Lin2017p}. 
We discuss in detail several models for the active force $\bm{F}_i^{(\rm active)}$ in Sec. \ref{sec:activity_polar} and \ref{sec:activity_nematic}. 

\textit{Dissipation.} -- 
We will consider three forces contributing to dissipation: 
\begin{align}
    \bm{F}_{i}^{\left( \text{dissipation} \right)} = \bm{F}_i^{(\rm friction)} + \bm{F}_{i}^{\left( \text{interface} \right)} + \bm{F}_{i}^{\left( \text{bulk} \right)},
\end{align}
each corresponding to the cell--substrate friction, the cell--cell interfacial viscosity, and the cell bulk viscosity, in turn. 

We consider a linear relation between the friction $\bm{F}_i^{(\rm friction)}$ and the vertex velocity $\bm{v}_i$: $\bm{F}_i^{\rm (friction)} = -{{\bm{\gamma} }_{i}}\cdot{{\bm{v}}_{i}}$ with ${{\bm{\gamma}}_{i}}$ being the 2D tensor friction coefficient between the vertex $i$ and the substrate. 
In general, ${{\bm{\gamma}}_{i}}$ can depend on cell shape and the substrate patterns. 
In the isotropic case, $\bm{\gamma}_i = \gamma_i \bm{I}$ with $\gamma_i$ being a scalar friction coefficient and $\bm{I}$ the second-order unit tensor. 

Considering the cell--cell interfacial viscosity and cell bulk viscosity as described in Sec. \ref{sec:cell_viscosity}, $\bm{F}_{i}^{\left( \text{interface} \right)}$ and $\bm{F}_{i}^{\left( \text{bulk} \right)}$ are related to the dissipation rates $S_{\rm interface}$ and $S_{\rm bulk}$ as: $\bm{F}_i^{(\rm interface)} = - \partial S_{\rm interface} / \partial \bm{v}_i$ and $\bm{F}_i^{\rm (bulk)} = - \partial S_{\rm bulk} / \partial \bm{v}_i$. 
Using the expressions of $S_{\rm interface}$ (Eq. \eqref{eq:Dissipation_Interface}) and $S_{\rm bulk}$ (Eq. \eqref{eq:Dissipation_Bulk}), we obtain: 
\begin{align}
\bm{F}_{i}^{\left( \text{interface} \right)} = & -\frac{\partial S_{\rm interface}}{\partial\bm{v}_i} \notag \\
= & \sum\limits_{j\in {{V}_{i}}}{\eta _{ij}^{\left( s \right)}\left( {{\bm{t}}_{i,j}}\otimes {{\bm{t}}_{i,j}} \right)}\cdot \left( {{\bm{v}}_{j}} - {{\bm{v}}_{i}} \right) , \label{eq:Force_InterfaceDissipation}
\end{align}
and 
\begin{align}
& \bm{F}_{i}^{\left( \text{bulk} \right)} = -\frac{\partial S_{\rm bulk}}{\partial \bm{v}_i} \notag \\ 
& = \sum\limits_{J\in {{C}_{i}}}{\left\{ \begin{aligned}
  & \eta_{J}^{\left( {b} \right)}\left( {{\bm{t}}_{i,J}}\otimes {{\bm{t}}_{i,J}} \right)\cdot \left( {{\bm{v}}_{J}}-{{\bm{v}}_{i}} \right) \\ 
 & -\frac{1}{{{n}_{J}}}\sum\limits_{k\in \cellJ}{\eta _{J}^{\left( {b} \right)}\left( {{\bm{t}}_{k,J}}\otimes {{\bm{t}}_{k,J}} \right)\cdot \left( {{\bm{v}}_{J}}-{{\bm{v}}_{k}} \right)} \\ 
\end{aligned} \right\}}
 , \label{eq:Force_BulkDissipation}
\end{align}
where $V_i$ and $C_i$ are the sets of neighboring vertices and neighboring cells of vertex $i$. 

The cell bulk dissipation force $\bm{F}_i^{\rm (bulk)}$ can be re-expressed as 
\begin{equation}
\bm{F}_{i}^{\left( \text{bulk} \right)} = \sum_{J \in C_i} \bm{F}_{J,i}^{\rm (bulk)} , 
\end{equation}
where 
\begin{align}
\bm{F}_{J,i}^{\rm (bulk)} = & \ \eta _{J}^{\left( {b} \right)}\left( {{\bm{t}}_{i,J}}\otimes {{\bm{t}}_{i,J}} \right)\cdot \left( {{\bm{v}}_{J}}-{{\bm{v}}_{i}} \right) \notag \\ 
 & -\frac{1}{{{n}_{J}}}\sum\limits_{k\in \cellJ}{\eta _{J}^{\left( {b} \right)}\left( {{\bm{t}}_{k,J}}\otimes {{\bm{t}}_{k,J}} \right)\cdot \left( {{\bm{v}}_{J}}-{{\bm{v}}_{k}} \right)} , \label{eq:Force_BulkDissipation_J}
\end{align}
represents the cell bulk dissipation force applied at vertex $i$ induced by its neighboring cell $J$. It is easy to prove that $\sum_{i \in \cellJ}{\bm{F}_{J,i}^{\rm (bulk)}} = \bm{0}$ and $\sum_{i \in \cellJ} \bm{r}_i \times \bm{F}_{J,i}^{\rm (bulk)} = \bm{0}$, indicating that the cell bulk dissipation forces $\bm{F}_{J,i}^{\rm (bulk)}$ are self-balancing. 
Thus, using the Batchelor formula \cite{Batchelor1970,Lau2009,NestorBergmann2018}, one can define a cellular bulk viscous stress as 
\begin{equation}
\bm{\sigma}_J^{\rm (bulk)} = -\frac{1}{A_J} \sum_{i \in \cellJ} \bm{r}_i \otimes \bm{F}_{J,i}^{\rm (bulk)} . \label{eq:Stress_BulkDissipation_0}
\end{equation}
Substituting Eq. \eqref{eq:Force_BulkDissipation_J} into Eq. \eqref{eq:Stress_BulkDissipation_0}, we obtain, 
\begin{align}
\bm{\sigma }_{J}^{\left( \rm bulk \right)} & =\frac{1}{{{A}_{J}}}\sum\limits_{i\in \cellJ}{\eta _{J}^{\left( {b} \right)}{{\ell }_{i,J}}\left[ {{\bm{t}}_{i,J}}\cdot \left( {{\bm{v}}_{J}}-{{\bm{v}}_{i}} \right) \right]{{\bm{t}}_{i,J}}\otimes {{\bm{t}}_{i,J}}} \notag \\ 
& = \frac{1}{{{A}_{J}}}\sum\limits_{i\in \cellJ}{\Lambda _{i,J}^{\left( \text{viscous} \right)}{{\ell }_{i,J}}{{\bm{t}}_{i,J}}\otimes {{\bm{t}}_{i,J}}} , 
\end{align}
where $\Lambda_{i,J}^{\rm (viscous)}$ is an effective viscous line tension of the virtual link between vertex $i$ and the geometric center of its neighboring cell $J$, as given by Eq. \eqref{eq:ViscousTension_Bulk}. 

\textit{Fluctuations.} -- 
The stochastic fluctuations can be implemented via an additional tension term, see Eqs. \eqref{eq:InterfacialTension} and \eqref{eq:BulkTension}.

\subsubsection{Matrix formalism of motion equation}

Equation \eqref{eq_ForceBalance} gives the motion equation of each vertex $i$. 
Here, we write it explicitly. Substituting Eqs. \eqref{eq:Force_InterfaceDissipation} and \eqref{eq:Force_BulkDissipation} into Eq. \eqref{eq_ForceBalance}, we obtain 
\begin{align}
  & \left[ \begin{aligned}
  & {{\bm{\gamma }}_{i}}+\sum\limits_{j\in {{V}_{i}}}{\eta _{ij}^{\left( {s} \right)}{{\bm{t}}_{i,j}}\otimes {{\bm{t}}_{i,j}}} \\ 
 & +\sum\limits_{J\in {{C}_{i}}}{\eta _{J}^{\left( {b} \right)}\left( \frac{{{n}_{J}}-2}{{{n}_{J}}}{{\bm{t}}_{i,J}}\otimes {{\bm{t}}_{i,J}}+\frac{1}{{{n}_{J}}}{{\bm{M}}_{J}} \right)} \\ 
\end{aligned} \right]\cdot {{\bm{v}}_{i}} \notag \\ 
 & -\sum\limits_{j\in {{V}_{i}}}{\eta _{ij}^{\left( {s} \right)}{{\bm{t}}_{i,j}}\otimes {{\bm{t}}_{i,j}}\cdot {{\bm{v}}_{j}}} \notag \\ 
 & -\sum\limits_{J\in {{C}_{i}}}{\frac{\eta _{J}^{\left( {b} \right)}}{{{n}_{J}}}\sum\limits_{\begin{smallmatrix} 
 j\in \cellJ \\ 
 j\ne i 
\end{smallmatrix}}{\left( {{\bm{t}}_{i,J}}\otimes {{\bm{t}}_{i,J}}+{{\bm{t}}_{j,J}}\otimes {{\bm{t}}_{j,J}}-{{\bm{M}}_{J}} \right)\cdot {{\bm{v}}_{j}}}} \notag \\
& = \bm{F}_{i}^{\left( \text{t} \right)} , \label{eq:ForceBalance_1}
\end{align}
where
\begin{equation}
{{\bm{M}}_{J}} = \frac{1}{{{n}_{J}}}\sum\limits_{k\in \cellJ}{{{\bm{t}}_{k,J}}\otimes {{\bm{t}}_{k,J}}} 
\end{equation}
is a tensor characterizing cell shape and $\bm{F}_{i}^{\left( \text{t} \right)}$ is the total non-dissipative vertex force that is independent of the vertices' velocities (including elastic forces, active forces, fluctuations, etc). 
The matrix form of Eq. \eqref{eq:ForceBalance_1} reads, 
\begin{equation}
\sum\limits_{j=1}^{N_v}{{{\bm{C}}_{ij}}\cdot {{\bm{v}}_{j}}}=\bm{F}_{i}^{\left( \text{t} \right)} , \label{eq:ForceBalance_MatrixForm_0}
\end{equation}
where 
\begin{equation}
{{\bm{C}}_{ij}}=\bm{C}_{ij}^{\left( {f} \right)}+\bm{C}_{ij}^{\left( s \right)}+\bm{C}_{ij}^{\left( b \right)} . 
\end{equation}
Here, $\bm{C}_{ij}^{\left( {f} \right)}$, $\bm{C}_{ij}^{\left( s \right)}$ and $\bm{C}_{ij}^{\left( b \right)}$ are the coefficient matrices of size $2 \times 2$, corresponding to the cell–substrate friction, the cell–cell interfacial viscosity and the cell bulk viscosity, with the detailed expressions given below:  
\begin{equation}
\bm{C}_{ij}^{\left( {f} \right)}={{\bm{\gamma}}_{i}}{{\delta }_{ij}}
\end{equation}
\begin{equation}
\bm{C}_{ij}^{\left( s \right)}=\left\{ \begin{aligned}
& \sum\limits_{k\in {{V}_{i}}}{\eta _{ik}^{\left( s \right)}\left( {{\bm{t}}_{i,k}}\otimes {{\bm{t}}_{i,k}} \right)}\ \ \ ,\ \ \ j=i \\ 
& -\eta _{ij}^{\left( s \right)}\left( {{\bm{t}}_{i,j}}\otimes {{\bm{t}}_{i,j}} \right)\ \ \ ,\ \ \ j\in {{V}_{i}} \\ 
& \bm{0}\ \ \ ,\ \ \ {\rm otherwise} \\ 
\end{aligned} \right. \label{eq:Cs_ij}
\end{equation}
\begin{equation}
\bm{C}_{ij}^{\left( b \right)}=\left\{ \begin{aligned}
  & \sum\limits_{J\in {{C}_{i}}}{\eta _{J}^{\left( b \right)}\left[
  \begin{aligned}
      & \frac{\left( {{n}_{J}}-2 \right)}{{{n}_{J}}}\left( {{\bm{t}}_{i,J}}\otimes {{\bm{t}}_{i,J}} \right) \\
      & + \frac{1}{n_J}\bm{M}_J
  \end{aligned}
  \right]} \ \ \ , \ \ \ j=i \\ 
 & \sum\limits_{i,j\in \cellJ}{\frac{\eta _{J}^{\left( {b} \right)}}{{{n}_{J}}}\left( \begin{aligned}
  & {{\bm{M}}_{J}}-{{\bm{t}}_{i,J}}\otimes {{\bm{t}}_{i,J}} \\ 
 & -{{\bm{t}}_{j,J}}\otimes {{\bm{t}}_{j,J}} \\ 
\end{aligned} \right)}
 \ \ \ ,\ \ \ j\ne i \\ 
 & \bm{0}\ \ \ ,\ \ \ \text{otherwise} \\ 
\end{aligned} \right. \label{eq:Cb_ij}
\end{equation}
where $\sum_{i,j \in \cellJ}$ sums over cells that contain both vertex $i$ and vertex $j$. 

Equation \eqref{eq:ForceBalance_MatrixForm_0} gives the force balance at each vertex $i$; in total, there are $N_v$ force balance equations. We can further express all the force balance equations in a unified matrix form, as shown in Eq. \eqref{eq:ForceBalance_MatrixForm}. 
Note that for the general case, the force vector $\bm{F}^{(\rm t)}$ is not limited to the forces demonstrated above, but instead can include all kinds of cell velocity-independent forces. $\bm{C}$, $\bm{v}$ and $\bm{F}_{{}}^{\left( \text{t} \right)}$ are organized by $\bm{C}_{ij}$, $\bm{v}_{i}$ and $\bm{F}_{i}^{(\rm t)}$ as follows: 
\begin{equation}
\bm{C}={{\left( {{\bm{C}}_{ij}} \right)}_{2N_v \times 2N_v}}={{\left( \begin{matrix}
   {{\bm{C}}_{11}} & \ldots  & {{\bm{C}}_{1 N_v}}  \\
   \vdots  & \ddots  & \vdots   \\
   {{\bm{C}}_{N_v 1}} & \cdots  & {{\bm{C}}_{N_v N_v}}  \\
\end{matrix} \right)}_{2N_v \times 2N_v}}
\end{equation}
\begin{equation}
\bm{v}={{\left( {{\bm{v}}_{i}} \right)}_{2N_v \times 1}}={{\left( \begin{aligned}
  & {{\bm{v}}_{1}} \\ 
 & \ \vdots  \\ 
 & {{\bm{v}}_{N_v}} \\ 
\end{aligned} \right)}_{2N_v \times 1}}
\end{equation}
\begin{equation}
\bm{F}_{{}}^{\left( \text{t} \right)}=\left( \bm{F}_{i}^{\left( \text{t} \right)} \right)_{2N_v \times 1} = {{\left( \begin{aligned}
  & \bm{F}_{1}^{\left( \text{t} \right)} \\ 
 & \ \vdots  \\ 
 & \bm{F}_{N_v}^{\left( \text{t} \right)} \\ 
\end{aligned} \right)}_{2N_v \times 1}}
\end{equation}
The friction-viscosity coefficient matrix $\bm{C}$ can be further decomposed as 
\begin{equation}
\bm{C}=\bm{C}_{{}}^{\left( {f} \right)}+\bm{C}_{{}}^{\left( s \right)}+\bm{C}_{{}}^{\left( b \right)} , 
\end{equation}
where $\bm{C}_{{}}^{\left( {f} \right)}$, $\bm{C}_{{}}^{\left( s \right)}$ and $\bm{C}_{{}}^{\left( b \right)}$ are the friction coefficient matrix, the cell–cell interfacial viscosity coefficient matrix and the cell bulk viscosity coefficient matrix: 
\begin{align}
{{\bm{C}}^{\left( {f} \right)}} & =\left( \bm{C}_{ij}^{\left( {f} \right)} \right)={{\left( \begin{matrix}
   \bm{C}_{11}^{\left( {f} \right)} & \ldots  & \bm{C}_{1 N_v}^{\left( {f} \right)}  \\
   \vdots  & \ddots  & \vdots   \\
   \bm{C}_{N_v 1}^{\left( {f} \right)} & \cdots  & \bm{C}_{N_v N_v}^{\left( {f} \right)}  \\
\end{matrix} \right)}_{2N_v \times 2N_v}} \notag \\
& ={{\left( \begin{matrix}
   {{\bm{\gamma}}_{1}} & \ldots  & \bm{0}  \\
   \vdots  & \ddots  & \vdots   \\
   \bm{0} & \cdots  & {{\bm{\gamma}}_{N_v}}  \\
\end{matrix} \right)}_{2N_v \times 2N_v}}
\end{align}
\begin{equation}
{{\bm{C}}^{\left( s \right)}}=\left( \bm{C}_{ij}^{\left( s \right)} \right)=\left( \begin{matrix}
   \bm{C}_{11}^{\left( s \right)} & \ldots  & \bm{C}_{1 N_v}^{\left( s \right)}  \\
   \vdots  & \ddots  & \vdots   \\
   \bm{C}_{N_v 1}^{\left( s \right)} & \cdots  & \bm{C}_{N_v N_v}^{\left( s \right)}  \\
\end{matrix} \right)_{2N_v \times 2N_v}
\end{equation}
\begin{equation}
{{\bm{C}}^{\left( b \right)}}=\left( \bm{C}_{ij}^{\left( b \right)} \right)=\left( \begin{matrix}
   \bm{C}_{11}^{\left( b \right)} & \ldots  & \bm{C}_{1 N_v}^{\left( b \right)}  \\
   \vdots  & \ddots  & \vdots   \\
   \bm{C}_{N_v 1}^{\left( b \right)} & \cdots  & \bm{C}_{N_v N_v}^{\left( b \right)}  \\
\end{matrix} \right)_{2N_v \times 2N_v}
\end{equation}

Note that, ${{\left[ \bm{C}_{ij}^{\left( s \right)} \right]}^{\text{T}}} = \bm{C}_{ij}^{\left( s \right)}=\bm{C}_{ji}^{\left( s \right)}$ and ${{\left[ \bm{C}_{ij}^{\left( b \right)} \right]}^{\text{T}}}=\bm{C}_{ij}^{\left( b \right)} = \bm{C}_{ji}^{\left( b \right)}$. 
Therefore, both the cell--cell interfacial viscosity coefficient matrix ${{\bm{C}}^{\left( s \right)}}$ and the cell bulk viscosity coefficient matrix ${{\bm{C}}^{\left( b \right)}}$ are symmetric matrices, resulting in a symmetric friction-viscosity coefficient matrix $\bm{C}$.

\subsection{Discussion}

In this Discussion section, we review previously proposed implementations of cellular viscosity, with a particular focus on the questions of (1) rotational invariance and (2) absence of friction on a fixed substrate.

\subsubsection{Our model is rotationally invariant}

Here we show that our model is invariant under a global rigid rotation represented by an orthogonal matrix $\bm{R}$ satisfying
$\bm{R}^\mathsf{T} \cdot \bm{R} = \bm{I}$ with $\bm{I}$ the identity matrix. 
Vertices' positions and velocities transform as
$\bm{r}_i' = \bm{R} \cdot \bm{r}_i$ and $\bm{v}_i' = \dot{\bm{r}}_i' = \bm{R}\cdot\dot{\bm{r}}_i + \dot{\bm{R}}\cdot\bm{r}_i = \bm{R}\cdot\bm{v}_i + \dot{\bm{R}}\cdot\bm{r}_i$. 
The unit vector
$\bm{t}_{i,j} = (\bm{r}_j - \bm{r}_i)/|\bm{r}_j - \bm{r}_i|$ then transforms as
\begin{equation}
\bm{t}_{i,j}' = 
\frac{\bm{r}_j' - \bm{r}_i'}{|\bm{r}_j' - \bm{r}_i'|}
= \frac{\bm{R}\cdot(\bm{r}_j - \bm{r}_i)}{|\bm{R}\cdot(\bm{r}_j - \bm{r}_i)|}
= \bm{R}\cdot\bm{t}_{i,j},
\end{equation}
where we have used the identity $|\bm{R}\cdot\bm{a}| = |\bm{a}|$ for any vector $\bm{a}$. Hence
\begin{align}
& \bm{t}_{i,j}'\cdot(\bm{v}_j' - \bm{v}_i') \notag \\
& =  (\bm{R}\cdot\bm{t}_{i,j})\cdot\left[\bm{R}\cdot(\bm{v}_j - \bm{v}_i) + \dot{\bm{R}}\cdot(\bm{r}_j-\bm{r}_i) \right] \notag \\
& = \bm{t}_{i,j}\cdot\bm{R}^{\mathsf{T}}\cdot\bm{R}\cdot(\bm{v}_j - \bm{v}_i) + \ell_{ij}\bm{t}_{i,j}\cdot\bm{R}^{\mathsf{T}}\cdot\dot{\bm{R}}\cdot\bm{t}_{i,j} \notag \\
& = \bm{t}_{i,j}\cdot(\bm{v}_j - \bm{v}_i) + \ell_{ij}\bm{t}_{i,j}\cdot\bm{\Omega}\cdot\bm{t}_{i,j} \notag \\
& = \bm{t}_{i,j}\cdot(\bm{v}_j - \bm{v}_i) , 
\end{align}
where $\bm{\Omega} = \bm{R}^{\mathsf{T}}\cdot\dot{\bm{R}}$ is a second-order antisymmetric tensor, thus satisfying $\bm{a}\cdot\bm{\Omega}\cdot\bm{a} = 0$ for any vector $\bm{a}$. 
Therefore, 
\begin{align}
{\Lambda_{i,j}^{(\rm viscous)}}'
 & = \eta_{ij}^{(s)} \bm{t}_{i,j}' \cdot (\bm{v}_j' - \bm{v}_i')
= \eta_{ij}^{(s)} \bm{t}_{i,j} \cdot (\bm{v}_j - \bm{v}_i) \notag \\
& = \Lambda_{i,j}^{(\rm viscous)} ,
\end{align}
showing that the viscous cell--cell interfacial tension defined in Eq. \eqref{eq:ViscousTension_Interface} is invariant under global rotations. 
In addition, it is easy to prove that the dissipation forces Eqs. \eqref{eq:Force_InterfaceDissipation} and \eqref{eq:Force_BulkDissipation} satisfy $\bm{F}_i^{\rm (interface)'} = \bm{R}\cdot\bm{F}_i^{\rm (interface)}$ and $\bm{F}_i^{\rm (bulk)'} = \bm{R}\cdot\bm{F}_i^{\rm (bulk)}$, further confirming that our viscous vertex model is rotationally invariant. 

The alternative formulation provided in previous studies (e.g., Refs. \cite{Okuda2015,Rozman2025}) is to assign a viscous force at vertex $i$ of the form
\begin{equation}
\bm{F}_i^{(\mathrm{visc,alt})}
= - \eta \sum_{j\in V_i} \bigl( \dot{\bm{r}}_i - \dot{\bm{r}}_j \bigr) . 
\label{eq:alternative}
\end{equation}
This expression is \emph{not} invariant under superposed rigid-body rotations (it violates material objectivity). 
Indeed, consider adding to the motion a global rigid rotation $\bm{R}$, the alternative viscous force Eq. (\ref{eq:alternative}) then transforms as
\begin{equation}
\bm{F}_i^{(\mathrm{visc,alt})'} = \bm{R}\cdot\bm{F}_i^{(\mathrm{visc,alt})}
- \eta \sum_{j\in V_i} \dot{\bm{R}}\cdot(\bm{r}_i - \bm{r}_j) . 
\end{equation}
The second term is generically nonzero; in this model, a pure rigid-body rotation (no edge stretching) leads to dissipation. 
Yet a junctional viscosity should not penalize global rotations. 
In contrast, the junctional viscous law used in our model, $\Lambda_{ij}^{(\mathrm{viscous})} \propto \bm{t}_{i,j}\cdot(\dot{\bm{r}}_j - \dot{\bm{r}}_i)$, is invariant under a global rigid rotation and therefore respects rotational symmetry.

\subsubsection{Thermodynamic stability and positive semi-definiteness of viscosity coefficient matrices}

To ensure that the friction-viscosity coefficient matrix $\bm{C} = \bm{C}^{(f)} + \bm{C}^{(s)} + \bm{C}^{(b)}$ represents a thermodynamically stable system where internal viscous forces do not perform positive work, $\bm{C}$ must be symmetric positive semi-definite ($\bm{C} \succeq 0$). 
Here, we analyze the interfacial and bulk viscosity contributions. 

Note that the cell--cell interfacial dissipation force $\bm{F}_i^{\rm (interface)}$ and the cell--cell interfacial viscosity coefficient matrix $\bm{C}_{ij}^{(s)}$ satisfy: 
\begin{equation}
\bm{F}_i^{\rm (interface)} = - \sum_{j=1}^{N_v} \bm{C}_{ij}^{(s)} \cdot \bm{v}_j , 
\end{equation}
which results in, 
\begin{equation}
\bm{C}_{ij}^{(s)} = - \frac{\partial \bm{F}_i^{(\rm interface)}}{\partial \bm{v}_j} = \frac{\partial^2 S_{\rm interface}}{\partial\bm{v}_i \partial\bm{v}_j} . 
\end{equation}
It suggests that the cell--cell interfacial viscosity coefficient matrix $\bm{C}^{(s)}$ corresponds to the Hessian matrix of the dissipation rate at cell--cell interfaces. 

Using the definition of $\bm{C}_{ij}^{(s)}$, it is easy to prove that 
\begin{equation}
\bm{v}^{\textnormal{T}} \bm{C}^{(s)} \bm{v} = \sum_{\langle i, j \rangle} \eta_{ij}^{(s)} \left[ \bm{t}_{i,j} \cdot (\bm{v}_i - \bm{v}_j) \right]^2 = 2 S_{\rm interface} . 
\end{equation}
Given that the interfacial viscosity coefficients are non-negative ($\eta_{ij}^{(s)} \ge 0$), this quadratic form is non-negative ($\ge 0$ for all $\bm{v}$). Consequently, $\bm{C}^{(s)}$ is a symmetric positive semi-definite matrix. 

Next, we analyze the bulk viscosity coefficient matrix $\bm{C}^{(b)}$. 
Similarly, we have
\begin{equation}
\bm{C}_{ij}^{(b)} = - \frac{\partial \bm{F}_i^{\rm (bulk)}}{\partial \bm{v}_j} = \frac{\partial^2 S_{\rm bulk}}{\partial\bm{v}_i \partial\bm{v}_j} , 
\end{equation}
suggesting that the cell bulk viscosity coefficient matrix $\bm{C}^{(b)}$ corresponds to the Hessian matrix of the dissipation rate at the cell bulk. Using the definition of $\bm{C}_{ij}^{(b)}$, we can prove
\begin{equation}
\bm{v}^{\textnormal{T}} \bm{C}^{(b)} \bm{v} = \sum_{J = 1}^{N_c}\sum_{i \in \cellJ} \eta_J^{(b)} [ \bm{t}_{i,J} \cdot (\bm{v}_J - \bm{v}_i) ]^2 = 2 S_{\rm bulk} . 
\end{equation}
Thus, given that $\eta_J^{(b)} \geq 0$, $\bm{C}^{(b)}$ is a symmetric positive semi-definite matrix. 

Therefore, given that all the cell viscosity coefficients are non-negative, i.e., $\eta_{ij}^{(s)} \geq 0$ and $\eta_J^{(b)} \geq 0$, the physical dissipation matrix $\bm{C} = \bm{C}^{(f)} + \bm{C}^{(s)} + \bm{C}^{(b)}$ is symmetric positive semi-definite.

\begin{figure}[t!]
\includegraphics[width=8.6cm]{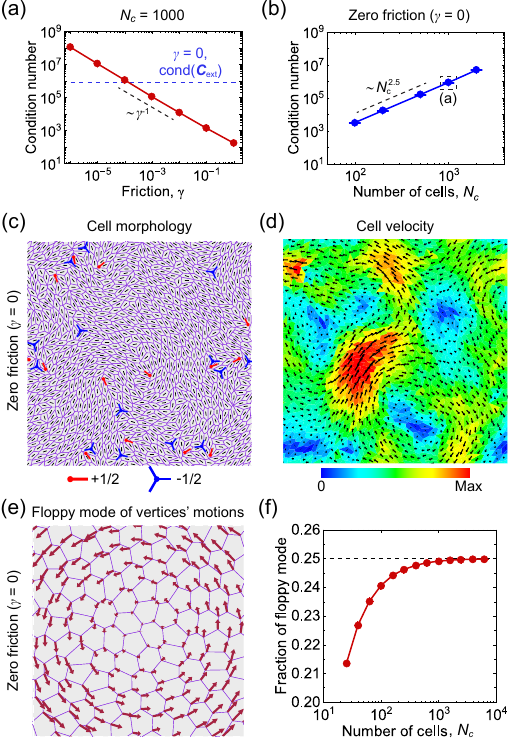}
\caption{\label{fig:zero_friction}
Numerical simulation of an active cell sheet in a square domain with periodic boundary conditions. 
Here, we consider cell-shape-dependent active stresses; see Sec. \ref{sec:activity_nematic}. 
(a) The condition number of the friction-viscosity coefficient matrix $\bm{C}$ as a function of the friction $\gamma$, for a system consisting of $N_c = 10^{3}$ cells. 
The dashed blue line represents the condition number of the extended friction-viscosity matrix $\bm{C}_{\rm ext}$ (see Eq. \eqref{eq:C_extended}). 
Data analysis gives a scaling law, ${\rm cond}(\bm{C}) \sim \gamma^{-1}$.
(b-d) The case without friction ($\gamma = 0$), using the protocol proposed in Sec. \ref{sec:zero-friction}. 
(b) The condition number of the extended friction-viscosity matrix $\bm{C}_{\rm ext}$ (see Eq. \eqref{eq:C_extended}) as a function of the number of cells $N_c$. 
Data analysis gives a scaling law, ${\rm cond}(\bm{C}_{\rm ext}) \sim N_c^{\alpha}$ with $\alpha \approx 2.5$. 
(c) A typical cell morphology. The black lines indicate cell orientations; the red (resp. blue) symbols represent $+1/2$ (resp. $-1/2$) topological defects, extracted using the scheme proposed in Ref. \cite{Lin2023}. 
(d) A typical flow field, where the black arrows represent the velocity vectors and the color code refers to the velocity magnitude. 
(e, f) Floppy modes of vertices' motions in the case of no friction ($\gamma = 0$) and no bulk viscosity ($\eta_b = 0$). 
(e) Illustration of a typical floppy mode in a cell layer consisting of $N_c = 100$ cells, obtained by numerical calculation of Eq. \eqref{eq:floppy_mode}. 
The arrows represent the vertices' velocity vectors. 
(f) The fraction of floppy modes $f$ as a function of the number of cells $N_c$. 
Parameters: $\eta_s = 10$, $T_0 = 0$, and $\beta = 0.4$; $\eta_b = 10$ for (a-d).}
\end{figure}

\subsubsection{Implementation of the zero-friction case}
\label{sec:zero-friction}

Here, we address in detail the limit of vanishing cell--substrate friction. 
This point is briefly addressed in the PhD thesis \cite{Staple2012Understanding} (p. 45), in which it is mentioned that, even in the presence of area, perimeter, and junctional viscosity, the eigenvalues of $\bm{C}$ could become numerically small when the friction to the substrate is considered negligible. We investigate this question and check that the limit of a vanishing substrate friction $\gamma = 0$ can be considered in our numerical simulation scheme. 

Based on the expressions of $\bm{C}_{ij}^{(s)}$ and $\bm{C}_{ij}^{(b)}$ (see Eqs. \eqref{eq:Cs_ij} and \eqref{eq:Cb_ij}), we note that
\begin{equation}
\sum\limits_{j=1}^{N_v}{\bm{C}_{ij}^{\left( s \right)}}=\bm{0} \quad , \quad \sum\limits_{j=1}^{N_v}{\bm{C}_{ij}^{\left( b \right)}}=\bm{0} . 
\end{equation}
These two equations indicate that both the cell--cell interfacial viscosity coefficient matrix ${{\bm{C}}^{\left( s \right)}}$ and the cell bulk viscosity coefficient matrix ${{\bm{C}}^{\left( b \right)}}$ are singular. 
Therefore, in the limiting case of zero friction, i.e., ${{\bm{\gamma}}_{i}}=\bm{0}$, the total friction-viscosity coefficient matrix $\bm{C}$ is singular. 
Physically, this is due to the unconstrained global translation and rotation of cells/vertices. 

To deal with this issue, one can assume a small friction, i.e., $\gamma / \eta \rightarrow 0$. However, a very small friction may result in a large condition number with Euclidean norm \cite{Golub2013}, estimated through the function \texttt{cond} in MATLAB \cite{Moler2004}, of the friction-viscosity coefficient matrix $\bm{C}$ (Fig. \ref{fig:zero_friction}(a)), leading to numerical divergence, while a finite friction cannot describe the case of zero friction well. 

Here, we propose an alternative numerical simulation method. In the case of zero friction, there is no momentum exchange and no angular momentum exchange between the cell sheet and the environment. Furthermore, assuming that initially the total momentum and total angular momentum of the cell sheet system are both zero, we have the following constraints on $\bm{v}_i$: 
\begin{equation}
\sum_{i = 1}^{N_v} \bm{v}_i = \bm{0} , \label{eq:zero_friction_constraint_1}
\end{equation}
\begin{equation}
\sum_{i = 1}^{N_v} \bm{r}_i \times \bm{v}_i = \bm{0} . \label{eq:zero_friction_constraint_2}
\end{equation}
These two constraints can be re-expressed as
\begin{equation}
\bm{D} \cdot \bm{v} = \bm{0} , \label{eq:zero_friction_constraint}
\end{equation}
where $\bm{D}$ is a constraint matrix of size $3 \times 2N_v$. 
Introducing the Lagrange multiplier $\bm{\lambda }$, the motion equation \eqref{eq:ForceBalance_MatrixForm} along with the constraint Eq. \eqref{eq:zero_friction_constraint} can be expressed together as 
\begin{equation}
\bm{C}_{\rm ext}\cdot \left( \begin{aligned}
  & \bm{v} \\ 
 & \bm{\lambda } \\ 
\end{aligned} \right) =
\left( \begin{aligned}
  & {{\bm{F}}^{\left( \text{t} \right)}} \\ 
 & {{\bm{0}}} \\ 
\end{aligned} \right) , 
\end{equation}
where
\begin{equation}
\bm{C}_{\rm ext} = \left( \begin{matrix}
   \bm{C} & {{\bm{D}}^{\text{T}}}  \\
   \bm{D} & \bm{0}  \\
\end{matrix} \right) \label{eq:C_extended}
\end{equation}
is referred to as an extended friction-viscosity matrix. 

\textit{Either substrate friction or bulk viscosity is needed.}
The extended friction-viscosity matrix defined in Eq. \eqref{eq:C_extended} is singular when $\gamma = 0$ and $\eta_{J}^{(b)} = 0$. In this case, the only remaining source of dissipation in $\bm{C}$ is the junctional viscosity $\eta_{ij}^{(s)}$, which penalizes changes in junctional lengths but does not generically oppose all possible vertex motions; in particular, there exist collective vertex velocity fields $\bm{v}$ that preserve all edge lengths and, consistent with the constraint $\bm{D}\cdot\bm{v}=0$, also preserve all cell areas. Such area- and length-preserving deformations correspond to floppy modes of the vertex network: they generate no viscous or frictional forces, so that one can find $(\bm{v},\bm{\lambda}) \neq \bm{0}$ satisfying
\begin{equation}
\bm{C}_{\rm ext}\cdot
\begin{pmatrix}
\bm{v}\\[2pt]
\bm{\lambda}
\end{pmatrix}
=
\bm{0} . \label{eq:floppy_mode}
\end{equation}

In Fig. \ref{fig:zero_friction}(e), we provide an example of such floppy modes in a cell layer consisting of $N_c = 100$ cells in a square domain with periodic boundary conditions. 
To quantify the proportion of floppy modes, we define the fraction of floppy modes as $f = 1 - {\rm rank}(\bm{C}_{\rm ext})/N_{\rm ext}$ with $N_{\rm ext} = 2 N_v + 3$ being the size of the extended friction-viscosity matrix $\bm{C}_{\rm ext}$. We find that the fraction of floppy modes $f$ increases with the number of cells $N_c$. When $N_c \rightarrow +\infty$, $f \simeq 1/4$ (Fig. \ref{fig:zero_friction}(f)), corresponding to one floppy mode per cell, ($f = N_c / (2 N_v) = N_c / (4 N_c) = 1/4$), consistent with Ref. \cite{Staple2012Understanding}, which implicitly assumes periodic boundary conditions. 

The existence of these zero-dissipation modes implies that the extended friction–viscosity coefficient matrix $\bm{C}_{\rm ext}$ is not invertible when $\gamma = 0$ and $\eta_{J}^{(b)} = 0$. Introducing either a finite substrate friction $\gamma > 0$ or a finite bulk viscosity $\eta_{J}^{(b)} > 0$ lifts these floppy modes, thereby regularizing $\bm{C}_{\rm ext}$ and restoring its invertibility. 

A convenient way to state the condition is the following. Because $\bm{C}$ is symmetric positive semi-definite, it may possess a non-trivial nullspace (denoted ${\rm ker}(\bm{C})$), which corresponds physically to the floppy modes of the dissipative operator, i.e., velocity fields that produce no viscous dissipation. 
The block matrix $\bm{C}_{\rm ext}$ is invertible only if the constraints encoded by $\bm{D}$ eliminate all these modes. 
More precisely, the standard result for semi-definite saddle-point systems states that $\bm{C}_{\rm ext}$ is non-singular if and only if $\bm{D}$ has full row rank and ${\rm ker}(\bm{C}) \cap {\rm ker}(\bm{D}) = \{\bm{0}\}$ \cite{Benzi_Golub_Liesen_2005}. 
In other words, the constraints must act injectively on ${\rm ker}(\bm{C})$: non-trivial floppy modes of $\bm{C}$ should not satisfy the constraints. 
This implies that the number of independent constraints must satisfy ${\rm rank}(\bm{D}) \geq {\rm dim}[{\rm ker}(\bm{C})]$. 

In particular, we note that $\bm{C}$ could decompose into several independent blocks, e.g., when simulating disjoint cellular aggregates. In that case, the constraint matrix $\bm{D}$ must remove the floppy modes of each block separately; physically, this means that the constraints encoded by $\bm{D}$ must eliminate all rigid translations and rotations of each connected component.

The discussion presented here contrasts with the one in Ref. \cite{Staple2012Understanding}, which deals with $\bm{C}$, and which concludes that, numerically, $\gamma$ should be chosen sufficiently small such that results are unchanged when $\gamma$ is further decreased, but sufficiently large such that the eigenvalues of the coefficient matrix are numerically nonzero. 

\textit{Maximum simulation size when $\gamma = 0$ is $N_c \approx 6 \times 10^5$.}
In agreement with this argument, numerically, we never encountered cases where $\bm{C}_{\rm ext}$ was not invertible when $\gamma = 0$ as long as $\eta_{J}^{(b)} > 0$. 
We check that the condition number of $\bm{C}_{\rm ext}$ (denoted ${\rm cond} (\bm{C}_{\rm ext})$) remains numerically tractable (Fig. \ref{fig:zero_friction}(a, b)), thus ensuring its suitability for subsequent simulations. 
For example, for a cell sheet consisting of $N_c = 1000$ cells, ${\rm cond} (\bm{C}_{\rm ext}) < 10^6$. We show an example of simulating an active cell sheet without friction in Fig. \ref{fig:zero_friction}(c, d).

The numerical stability of the proposed framework is verified by analyzing the condition number of the friction-viscosity coefficient matrix (denoted ${\rm cond}(\bm{C})$) and that of the augmented saddle-point system (denoted ${\rm cond}(\bm{C}_{\text{ext}})$). 
As illustrated in Fig. \ref{fig:zero_friction}(a), ${\rm cond}(\bm{C})$ follows an inverse scaling with friction (${\rm cond} \propto \gamma^{-1}$); for a reference case of $N_c = 1000$ and $\gamma = 10^{-6}$, it reaches approximately $10^8$. 
Furthermore, the scaling with system size follows a power law ${\rm cond}(\bm{C}_{\rm ext}) \propto N_c^{\alpha}$ with $\alpha \approx 2.5$ (Fig. \ref{fig:zero_friction}(b)).

This scaling defines the framework's operational limits in terms of an overall number of cells around $N_c \approx 6 \times 10^5$, at which we expect ${\rm cond}(\bm{C}_{\rm ext})$ to exceed the solver capacity at $10^{13}$ \cite{Moler2004}, beyond which the relative error in the solution—governed by the product of the condition number and the machine epsilon of 64-bit double precision ($\sim 10^{-16}$) reaches $10^{-3}$. Beyond this point, numerical rounding errors begin to contaminate the first three significant digits of the velocity field, potentially obscuring the physical dynamics. For the tissue sizes considered in this study ($N_c \le 2000 \ll 10^6$), the system remains several orders of magnitude away from this unstable regime, ensuring that the kinematic constraints effectively regularize the vanishing-friction limit.

\textit{Boundary conditions can be redundant.}
Note that if fixed boundaries or externally applied forces exist, implementing these boundary conditions via the Lagrange multiplier method (see Sec. \ref{sec:tissue_viscosity} and \ref{sec:activity}) will also address the singularity issue of the friction-viscosity coefficient matrix. In such cases, the boundary constraints Eqs. \eqref{eq:zero_friction_constraint_1} and \eqref{eq:zero_friction_constraint_2} are not necessary.

\subsubsection{Alternative formulations of the viscous dissipation}
\label{sec:alternative_viscous_tension}

\textit{Alternative interfacial viscosity model.} -- 
We propose an alternative cell--cell interfacial viscous tension: 
\begin{equation}
{{\Lambda }_{ij}^{\rm (viscous)}} = \eta_{ij}^{(s)} \dot{\varepsilon}_{ij} , \label{eq:ViscousTension_Interface_2}
\end{equation}
where $\varepsilon_{ij}$ quantifies the elongation/shrinkage strain of the cell--cell interface $ij$. 
Since cell--cell interfaces can undergo large elongation or shrinkage, we employ the logarithmic strain to quantify the elongation/shrinkage of cell--cell interfaces, i.e., 
\begin{equation}
\varepsilon_{ij} = \ln \left( \frac{\ell_{ij}}{\ell_{ij,0}} \right) , \label{eq:strain_interface}
\end{equation}
where $\ell_{ij,0}$ refers to a rest-length parameter of the cell--cell interface $ij$. Such a formulation remains rotationally invariant. 
From Eq. \eqref{eq:strain_interface}, we obtain that $\dot{\varepsilon}_{ij} = {{{{\dot{\ell}}}_{ij}}}/{{{\ell}_{ij}}}$. Then, Eq. \eqref{eq:ViscousTension_Interface_2} further reads, 
\begin{equation}
{{\Lambda }_{ij}^{\rm (viscous)}} = \eta _{ij}^{\left( s \right)}\frac{{{{\dot{\ell}}}_{ij}}}{{{\ell}_{ij}}} . \label{eq:InterfacialViscousTension_3}
\end{equation}
Thus, the total cell--cell interfacial tension reads, ${{\Lambda }_{ij}} = \eta _{ij}^{( s )}{{{{\dot{\ell}}}_{ij}}}/{{{\ell}_{ij}}} + \Lambda _{ij}^{\left( 0 \right)} + {{\zeta }_{ij}} / {\sqrt{{{\ell}_{ij}}}}$.
Examining the cell--cell interfacial viscosity term, it suggests that if any cell--cell interface $ij$ becomes very short, i.e., $\ell_{ij} \rightarrow 0$ and $\dot{\ell}_{ij} < 0$, which decreases the cell--cell interfacial tension and even leads to a negative value, the shrinkage of the cell--cell interface will thus be resisted. This can be further directly illustrated by considering the 1D case. Setting $\Lambda_{ij} = 0$ and ignoring stochastic fluctuations, we obtain, $\dot{\ell}_{ij} = -{\Lambda_{ij}^{(0)}}\ell_{ij} / {\eta_{ij}^{(s)}}$, which leads to, $\ell_{ij}(t) = \ell_{ij,0}\exp [-({\Lambda_{ij}^{(0)}}/{\eta_{ij}^{(s)}})t ]$, where $\ell_{ij,0} = \ell_{ij}(t=0)$ here. 
It clearly shows that an increase in the cell--cell interfacial viscosity $\eta_{ij}^{(s)}$ slows the shrinkage of the cell--cell interface. Therefore, the cell--cell interfacial viscosity defined in Eq. \eqref{eq:ViscousTension_Interface_2} tends to suppress T1 topological transitions, that is, cell neighbor exchange \cite{Fletcher2014,Lin2018}. 
In comparison, with the definition of Eq. \eqref{eq:ViscousTension_Interface}, one can obtain $\ell_{ij}(t) = \ell_{ij,0} - [\Lambda_{ij}^{(0)}/\eta_{ij}^{(s)}]t$, which results in a finite shrinkage time of the cell--cell interface.
In this study, we adopt the definition in Eq. \eqref{eq:ViscousTension_Interface} rather than Eq. \eqref{eq:ViscousTension_Interface_2} for the viscous tension. 

\textit{Alternative bulk viscosity model.} -- 
Inspired by Ref. \cite{Brodland2009}, as well as motivated by recent observations of radial actin fibers --- termed ``actin stars'' in Ref. \cite{Barai2025} --- we consider here a dissipation mechanism along the axis connecting vertices to the cell center \cite{Barai2025}. 
Cellular viscosities were introduced by Staple \cite{Staple2012Understanding} via dissipative friction forces acting on the cell area, cell perimeter, and cell--cell interface length, expressed as:
\begin{equation}
\bm{F}_i^{\rm (viscous)} = - \sum_{m=1}^{N_m} \eta_m \dot{g}_m \frac{\partial g_m}{\partial \bm{r}_i}
\end{equation}
where $g_m$ represents the cell area, cell perimeter, or cell--cell interface length. 
Similarly, in the framework of Nestor-Bergmann \textit{et al.} \cite{NestorBergmann2018}, they introduced cell viscosities by constructing a cell stress tensor with viscous contributions proportional to the cell area expansion rate $\dot{A}_J$ (associated with a bulk viscosity $\mu_b$) and the cell perimeter growth rate $\dot{P}_J$ (associated with a cortical viscosity $\mu_c$).

\subsubsection{Model simplification} 
\label{sec:defaultparametervalues}

The viscous vertex model proposed here can be applied to account for spatial inhomogeneities in cell--substrate friction, in cell--cell interfacial viscosity, and in cell bulk viscosity. 
In the present study, for simplicity, we assume a homogeneous and isotropic friction, i.e., $\bm{\gamma}_i = \gamma\bm{I}$ with $\gamma$ being a scalar friction coefficient. 
We also assume a homogeneous cell--cell interfacial viscosity and a homogeneous cell bulk viscosity across the whole tissue. Thus, we set $\eta_{ij}^{(s)} = \eta_s$ for all cell--cell interfaces and $\eta_J^{(b)} = \eta_b$ for all cells. 
We further set the constant parts of line tensions to be zero, i.e., $\Lambda_{ij}^{(0)} = 0$ and $\Lambda_{iJ}^{(0)} = 0$. 

Note that the presence of stochastic tension fluctuations ($\zeta_{ij}$ and $\zeta_{i,J}$), when strong enough, can lead to spontaneous T1 topological transitions, thus fluidifying tissues. 
In our present study, we focus on the role of cellular viscosities and do not consider the tension fluctuations by setting $\zeta_{ij} = 0$ for all cell--cell interfaces $ij$ and $\zeta_{i,J} = 0$ for all vertex--cell links $iJ$ hereafter. 

In our numerical simulations, if not stated otherwise, we set the default parameter values as follows: $K_A = 1$, $A_0 = 1$, $K_P = 0.02$, $P_0 = 1$, and $\gamma = 1$.

\subsubsection{Simulation cost}

Compared to the purely frictional vertex model, accounting for cellular viscosity increases the computational cost of solving the algebraic equations in Eq. \eqref{eq:ForceBalance_MatrixForm}. This is due to the presence of off-diagonal components in the friction-viscosity coefficient matrix $\bm{C}$. For a cell sheet consisting of $N_c = 1000$ cells with $\gamma = 1$, the average CPU time required to solve Eq. \eqref{eq:ForceBalance_MatrixForm} is $\sim 2 \times 10^{-5}$\,s when $\eta_s = \eta_b = 0$, but rises to $\sim 5 \times 10^{-2}$\,s when $\eta_s = \eta_b = 1$ (tested on an Intel(R) Core(TM) Ultra 7, MATLAB R2024b). Furthermore, increased cellular viscosity leads to slower cellular motion; consequently, higher viscosity values require longer simulation times to reach a dynamic steady state.

\begin{figure*}[t!]
\includegraphics[width=17.5cm]{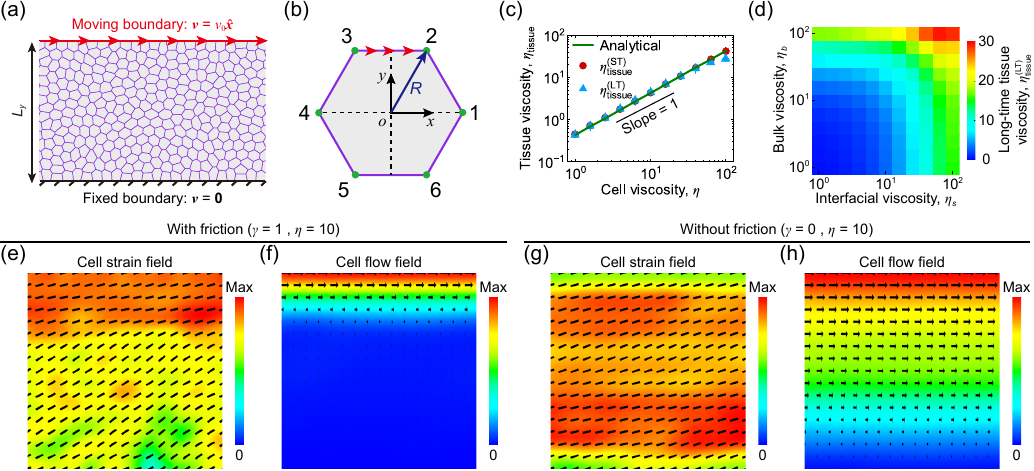}
\caption{\label{fig:shear_slab} 
Measuring the coarse-grained tissue viscosity. Here, we do not consider cell activities (i.e., $T_0 = 0$ and $\beta = 0$). 
(a) Schematic of shearing a cell sheet within a slab geometry. The slab is assumed to be periodic along the $x$-axis and is of width $L_y$ in the $y$-axis. Cell vertices adhered to the bottom border are fixed, i.e. $\bm{v} = \bm{0}$, while cell vertices adhered to the top border move at a specified velocity $\bm{v} = v_0 \hat{\bm{x}}$. 
Thus, the global tissue shear strain rate is $\dot{\gamma}_{xy} = v_0 / L_y$. 
(b) Theoretical analysis of a single hexagonal cell under a constant shear strain rate $\dot{\gamma}_{xy}$. 
(c) The short-time tissue viscosity $\eta_{\rm tissue}^{(\rm ST)}$ and the long-time tissue viscosity $\eta_{\rm tissue}^{\rm (LT)}$ as functions of cellular viscosity $\eta = \eta_s = \eta_b$, with $\gamma = 0$. 
Comparison with the analytical result (Eq. \eqref{eq:eta_tissue_analytical}). 
(d) The long-time tissue viscosity $\eta_{\rm tissue}^{\rm (LT)}$ regulated by the cellular viscosities ($\eta_s$, $\eta_b$). 
(e-h) Comparisons of the cell deformation strain field (e, g) and the cell flow field (f, h) for the case with friction (e, f) and the case without friction (g, h), averaged over $N_t = 100$ frames. 
In (e, g), the color code refers to the magnitude of the cell deformation tensor $\bm{\varepsilon}_{\rm cell}^{(\rm dev)}$ (the deviatoric part) and lines represent orientations. 
In (f, h), the color code indicates the magnitude of velocity, and the arrows represent the velocity vectors. Parameters are indicated in Sec. \ref{sec:defaultparametervalues}. 
}
\end{figure*}

\section{Coarse-grained tissue viscosity}
\label{sec:tissue_viscosity}

Here, we employ our viscous vertex model to correlate the coarse-grained tissue viscosity $\eta_{\rm tissue}$ with the interfacial and bulk cell viscosities $\eta_s$ and $\eta_b$ in a purely passive vertex model. 

To do so, we consider a tissue in a slab geometry. We apply a fixed shear strain rate $\dot{\gamma}_{xy}$ and measure the exerted stress.

\subsection{Method}

In this section, we consider a cell sheet adhered to a slab geometry of width $L_y$, as in Refs. \cite{Fu2024,Fang2022,Lin2019}, as shown in Fig. \ref{fig:shear_slab}(a). 
We apply a shear strain rate $\dot{\gamma}_{xy}$ to the cell sheet in the following way: (i) the cell vertices adhered to the bottom border are fixed, i.e., $\bm{v}_i = \bm{0}$ for all bottom vertices; (ii) the cell vertices adhered to the top border move at a prescribed velocity $\bm{v}_i = v_0 \hat{\bm{x}}$ for all top vertices with $v_0 = L_y\dot{\gamma}_{xy}$ and $\hat{\bm{x}}$ a unit vector along the long axis of the slab (i.e., $x$ axis). Therefore, we have the following boundary conditions: 
\begin{equation}
\bm{v}_i = 
\begin{cases}
\bm{0} & \text{bottom boundary} \\
v_0 \hat{\bm{x}} & \text{top boundary} 
\end{cases}
\end{equation}
which can be formulated in the matrix form as follows: 
\begin{equation} \label{eq:boundary}
\bm{D} \cdot \bm{v} = \bm{v}_{b} . 
\end{equation}
Introducing the Lagrange multiplier $\bm{\lambda }$, the vertices' velocities satisfy the following equation: 
\begin{equation}
\left( \begin{matrix}
   \bm{C} & {{\bm{D}}^{\text{T}}}  \\
   \bm{D} & \bm{0}  \\
\end{matrix} \right)\cdot \left( \begin{aligned}
  & \bm{v} \\ 
 & \bm{\lambda } \\ 
\end{aligned} \right)=\left( \begin{aligned}
  & {{\bm{F}}^{\left( \text{t} \right)}} \\ 
 & {{\bm{v}}_{b}} \\ 
\end{aligned} \right) \label{eq:MotionEquation_SlabShear}
\end{equation}
where the matrix $\bm{D}$ is defined through Eq. (\ref{eq:boundary}).

\subsection{Results} 
\subsubsection{Short-time tissue viscosity}

Given the configuration of a cell sheet, i.e., the vertices' positions $\{ \bm{r}_i \}$ and the network topology, we can define the short-time tissue viscosity $\eta_{\rm tissue}^{(\rm ST)}$ as follows. 

Solving Eq. \eqref{eq:MotionEquation_SlabShear}, we obtain the velocity vector $\bm{v}_i$ of each vertex $i$. Subsequently, we can calculate the viscous cell--cell interfacial tension $\Lambda_{ij}^{\rm (viscous)}$ via Eq. \eqref{eq:ViscousTension_Interface} and the viscous cell--bulk line tension $\Lambda_{i,J}^{\rm (viscous)}$ via Eq. \eqref{eq:ViscousTension_Bulk}. 
Then we can calculate the viscous shear stress of the tissue as 
\begin{align}
\sigma _{xy}^{\left( \text{viscous} \right)} = & \frac{\sum\limits_{<i,j>}{\Lambda _{ij}^{\left( \text{viscous} \right)}{{\ell}_{ij}}t_{ij}^{\left( x \right)}t_{ij}^{\left( y \right)}}}{\sum\limits_{J = 1}^{N_c}{{{A}_{J}}}} \notag \\
& + \frac{\sum\limits_{J=1}^{N_c}{\sum\limits_{i \in \ \text{cell}\ J}{\Lambda _{i,J}^{\left( \text{viscous} \right)}{{\ell}_{i,J}}t_{i,J}^{\left( x \right)}t_{i,J}^{\left( y \right)}}}}{\sum\limits_{J=1}^{N_c}{{{A}_{J}}}} . \label{eq_ViscousShearStress}
\end{align}
Finally, we can measure the coarse-grained short-time tissue viscosity as 
\begin{equation}
\eta_{\rm tissue}^{(\rm ST)} = \frac{\sigma_{xy}^{(\rm viscous)}}{\dot{\gamma}_{xy}} . 
\end{equation}
From Eqs. \eqref{eq:MotionEquation_SlabShear}--\eqref{eq_ViscousShearStress}, we know that the tissue viscosity $\eta_{\rm tissue}^{\rm (ST)}$ depends only on the friction-viscosity coefficient matrix $\bm{C}$, which depends on the cell--cell interfacial viscosity $\eta_s$, the cell bulk viscosity $\eta_b$ as well as the geometry (vertices' positions) and topology (neighboring relationship) of the cell sheet. However, it should be noted that the geometry and topology of the cell sheet can be affected by cell mechanical properties and cell activity, which thus indirectly affect the tissue viscosity. 

In addition, upon varying the shear strain rate $\dot{\gamma}_{xy}$, Eq. \eqref{eq:MotionEquation_SlabShear} becomes
\begin{equation}
\left( \begin{matrix}
   \bm{C} & {{\bm{D}}^{\text{T}}}  \\
   \bm{D} & \bm{0}  \\
\end{matrix} \right)\cdot \left( \begin{aligned}
  & \delta\bm{v} \\ 
 & \delta\bm{\lambda } \\ 
\end{aligned} \right)=\left( \begin{aligned}
  & \bm{0} \\ 
 & \delta{{\bm{v}}_{b}} \\ 
\end{aligned} \right) \label{eq:MotionEquation_SlabShear_Variation}
\end{equation}
Thus, we have $\delta\bm{v} \propto \delta\bm{v}_{b} \propto \delta\dot{\gamma}_{xy}$, which leads to $\delta\Lambda_{ij}^{(\rm viscous)} \propto \delta\dot{\gamma}_{xy}$ and $\delta\Lambda_{i,J}^{(\rm viscous)} \propto \delta\dot{\gamma}_{xy}$. Consequently, $\delta\sigma_{xy}^{(\rm viscous)} \propto \delta\dot{\gamma}_{xy}$. This suggests that the viscous shear stress $\sigma_{xy}^{(\rm viscous)}$ is linearly related to the shear rate $\dot{\gamma}_{xy}$, as validated by our numerical simulations. 

Our numerical simulations show that, in the absence of cell activity, the short-time tissue viscosity $\eta_{\rm tissue}^{(\rm ST)}$ is linearly related to the cell viscosity $\eta = \eta_s = \eta_b$ (Fig. \ref{fig:shear_slab}(c)). 
This is because, in the absence of cell activity, the cell viscosity does not affect the tissue morphology and topology. 

To gain an analytical expression of the short-time tissue viscosity, here, we consider a hexagonal cell of radius $R$, as shown in Fig. \ref{fig:shear_slab}(b). 
The coordinates of the $k$-th vertex read: $x_k = R \cos [(k-1)\pi/3]$ and $y_k = R \sin [(k-1)\pi/3]$ with $k = 1,2,\cdots,6$.  
Now we fix the fifth and sixth vertices and apply a uniform shear strain rate $\dot{\gamma}_{xy}$ to the hexagonal cell. The velocity of each vertex reads 
\begin{equation}
\begin{aligned}
& \bm{v}_1 = \bm{v}_4 = \frac{\sqrt{3}}{2}R\dot{\gamma}_{xy} \hat{\bm{x}} , \\ 
& \bm{v}_2 = \bm{v}_3 = \sqrt{3}R\dot{\gamma}_{xy} \hat{\bm{x}} , \\ 
& \bm{v}_5 = \bm{v}_6 = \bm{0} . \\ 
\end{aligned}
\end{equation}
Thus, we can calculate the viscous cell--cell interfacial tension as 
\begin{equation}
\begin{aligned}
& \Lambda _{12}^{\left( \text{viscous} \right)} = \Lambda _{45}^{\left( \text{viscous} \right)} = -\frac{\sqrt{3}}{4}{{\eta }_{s}}R{{{\dot{\gamma }}}_{xy}} , \\ 
 & \Lambda _{34}^{\left( \text{viscous} \right)} = \Lambda _{61}^{\left( \text{viscous} \right)} = \frac{\sqrt{3}}{4}{{\eta}_{s}}R{{{\dot{\gamma }}}_{xy}} , \\ 
 & \Lambda _{23}^{\left( \text{viscous} \right)} = \Lambda _{56}^{\left( \text{viscous} \right)} = 0 . \\ 
\end{aligned}
\end{equation}
and the viscous cell--bulk line tension as 
\begin{equation}
\begin{aligned}
& \Lambda_{1,J}^{\left( \text{viscous} \right)} = \Lambda _{4,J}^{\left( \text{viscous} \right)} = 0 , \\ 
& \Lambda_{2,J}^{\left( \text{viscous} \right)} = \Lambda_{5,J}^{\left( \text{viscous} \right)}=\frac{\sqrt{3}}{4}{{\eta }_{b}}R{{{\dot{\gamma }}}_{xy}} , \\
& \Lambda_{3,J}^{\left( \text{viscous} \right)} = \Lambda _{6,J}^{\left( \text{viscous} \right)} = -\frac{\sqrt{3}}{4}{{\eta }_{b}}R{{{\dot{\gamma }}}_{xy}} . \\ 
\end{aligned}
\end{equation}
Consequently, the total viscous shear stress reads
\begin{equation}
\sigma _{xy}^{\left( \text{viscous} \right)}=\frac{1}{4\sqrt{3}}{{\eta}_{s}}{{\dot{\gamma }}_{xy}}+\frac{1}{2\sqrt{3}}{{\eta}_{b}}{{\dot{\gamma }}_{xy}} . 
\end{equation}
Finally, we obtain the short-time tissue viscosity as 
\begin{equation} \label{eq:etaST}
\eta_{\rm tissue}^{\rm (ST)} = \frac{\sigma_{xy}^{\rm (viscous)}}{\dot{\gamma}_{xy}} = \frac{1}{4\sqrt{3}}{{\eta }_{s}} + \frac{1}{2\sqrt{3}}{{\eta }_{b}} . 
\end{equation}
In particular, when $\eta_s = \eta_b = \eta$, we obtain a simple relation between the short-time tissue viscosity and the cellular viscosity as 
\begin{equation}
\eta_{\rm tissue}^{\rm (ST)} = \frac{\sqrt{3}}{4}{\eta} . \label{eq:eta_tissue_analytical}
\end{equation}
This analytical expression agrees well with our numerical calculations (Fig. \ref{fig:shear_slab}(c)).

\subsubsection{Long-time tissue viscosity}

The calculation of the short-time tissue viscosity does not include cell--cell rearrangements, and can fail to account for the long-time tissue response to applied shear strain/stress.

Here, we apply a constant shear strain rate $\dot{\gamma}_{xy}$ to the tissue. 
We allow cell motion and cell--cell rearrangement and monitor the evolution of the tissue viscous stress $\sigma_{xy}^{(\rm viscous)} (t)$ under the sustained shear strain rate $\dot{\gamma}_{xy}$. 
This allows us to evaluate the long-time tissue viscosity as: 
\begin{equation}
\eta_{\rm tissue}^{(\rm LT)} = \frac{\langle \sigma_{xy}^{(\rm viscous)} (t) \rangle}{\dot{\gamma}_{xy}} , 
\end{equation}
where $\langle \cdot \rangle$ is an average over time. 

In Fig. \ref{fig:shear_slab}(d), we represent a phase diagram of the long-time tissue viscosity $\eta_{\rm tissue}^{(\rm LT)}$ as regulated by the two different cell viscosities $\eta_s$ and $\eta_b$. We find that $\eta_{\rm tissue}^{(\rm LT)}$ increases with both $\eta_s$ and $\eta_b$, and exhibits a linear scaling at small cell viscosity (Fig. \ref{fig:shear_slab}(c)). We also note that such a behavior is consistent with that of the short-time tissue viscosity $\eta_{\rm tissue}^{\rm (ST)}$, see Eq. (\ref{eq:etaST}). 

When friction dominates, we observe localized shear banding near the moving boundary (Fig. \ref{fig:shear_slab}(e,f)). 
As seen in Fig. \ref{fig:shear_slab}(e), the presence of substrate friction ($\gamma = 1$) leads to a sharp localization of the velocity field at the driven boundary. This profile exhibits a kink where the flow abruptly drops to zero: this is a hallmark of shear banding. A recent preprint explores this phenomenon in more depth \cite{nicholas2026fluidsolidpatternformationstrain}. 

When friction is negligible, i.e., $\gamma = 0$, as expected, shear banding vanishes, and the shear flow field is consistent with that observed in a Newtonian fluid, see Fig. \ref{fig:shear_slab}(g,h). 
In this regime, we find that the long-time tissue viscosity $\eta_{\rm tissue}^{(\rm LT)}$ is close to (but lower than) the short-time tissue viscosity $\eta_{\rm tissue}^{\rm (ST)}$, especially at small cellular viscosity. 
At larger cellular viscosities (larger $\eta_s$ and larger $\eta_b$), the long-time tissue viscosity $\eta_{\rm tissue}^{(\rm LT)}$ becomes distinguishable from the short-time tissue viscosity $\eta_{\rm tissue}^{\rm (ST)}$. 
This is because at high cellular viscosity, cell deformation cannot be relaxed on the timescale set by the applied shear. 
In particular, we check that when T1 topological transition is inhibited, the extracted long-time tissue viscosity will be elevated and even much larger than the short-time tissue viscosity.

\subsection{Discussion}

In Ref. \cite{NestorBergmann2018}, the coarse-grained tissue shear stress is obtained by summing the contributions of the cell area expansion rate and the cell perimeter growth rate over all cells; in the linear regime, the effective shear viscosity emerges as a geometric prefactor multiplying a strictly linear combination of microscopic viscosities. 

While a significant body of work has explored the long-time macroscopic tissue rheology of the vertex model \cite{Duclut2021,Tong2022,Grossman2025}, these studies generally focused on cases where substrate friction is the sole source of dissipation. Conversely, Tong \textit{et al.} \cite{Tong2023} developed a normal-mode analysis for vertex models incorporating both external (cell--substrate) and internal (cell--cell) dissipations, resulting in a macroscopic tissue viscosity that depends linearly on the microscopic ones --- a result consistent with ours in the short-time limit. However, cell rearrangements were not considered in Ref. \cite{Tong2023}.

Recently posted preprints have addressed related problems concerning the rheology of the vertex model with internal viscous dissipation. Anand \textit{et al.} \cite{Anand2026} explored the non-linear visco-elasto-plastic rheology of cell sheets, linking cell-level and tissue-level mechanics through mean-field rheological relations. Additionally, Nguyen \textit{et al.} \cite{nguyen2026cellcelladhesiondoubleedgedsword} proposed a microscopic model of viscous cell--cell dissipation and its impact on overall rheology, while Nicholas \textit{et al.} \cite{nicholas2026fluidsolidpatternformationstrain} discussed how internal dissipation can suppress shear banding that otherwise occurs under external substrate drag.

\section{Cell--cell dissipation and activity: large-scale tissue flows}
\label{sec:activity}

\subsection{Drag on a cell or a cell cluster}

To better illustrate the effect of cell viscosity, here, we perform numerical simulations of pulling a cell within a passive cell sheet. We first provide details on the method before discussing our results. 

\subsubsection{Method}

Let us first consider the case where a sustained drag is applied to a cell (denoted $J$) with a constant velocity ${{\bm{v}}_{0}}$. We then have the following boundary condition: 
\begin{equation}
\frac{1}{{{n}_{J}}}\sum\limits_{i\in \cellJ}{{{\bm{v}}_{i}}}={{\bm{v}}_{0}} , 
\end{equation}
that is 
\begin{equation}
\bm{D}\cdot \bm{v}={{\bm{v}}_{0}} , \label{eq_BoundaryConstraints_MatrixForm}
\end{equation}
where $\bm{D}={{\left( {{\bm{D}}_{i}} \right)}_{2\times 2N_v}}$ is the constraint matrix with $\bm{D}_i$ being 
\begin{equation}
{{\bm{D}}_{i}}=\left\{ \begin{aligned}
  \frac{1}{{{n}_{J}}}\bm{I}\ \ \  & ,\ \ \ i\in \cellJ \\ 
 \bm{0} \ \ \ \ \ & ,\ \ \ \text{otherwise} \\ 
\end{aligned} \right.
\end{equation}
Introducing the Lagrange multiplier $\bm{\lambda }={{\left( {{\lambda }_{1}},{{\lambda }_{2}} \right)}^{\text{T}}}$, the motion equation \eqref{eq:ForceBalance_MatrixForm} along with the boundary constraint Eq. \eqref{eq_BoundaryConstraints_MatrixForm} can be expressed together as 
\begin{equation}
\left( \begin{matrix}
   \bm{C} & {{\bm{D}}^{\text{T}}}  \\
   \bm{D} & \bm{0}  \\
\end{matrix} \right)\cdot \left( \begin{aligned}
  & \bm{v} \\ 
 & \bm{\lambda } \\ 
\end{aligned} \right)=\left( \begin{aligned}
  & {{\bm{F}}^{\left( \text{t} \right)}} \\ 
 & {{\bm{v}}_{0}} \\ 
\end{aligned} \right) , \label{eq_MotionEquation}
\end{equation}
which is equivalent to 
\begin{equation}
\left\{ \begin{aligned}
  & \bm{C}\cdot \bm{v}={{\bm{F}}^{\left( \text{t} \right)}}-{{\bm{D}}^{T}}\cdot \bm{\lambda } \\ 
 & \bm{D}\cdot \bm{v}={{\bm{v}}_{0}} \\ 
\end{aligned} \right.
\end{equation}
Therefore, the term $-{{\bm{D}}^{T}}\cdot \bm{\lambda }$ corresponds to the reaction force induced by the boundary constraint, i.e., the pulling force in the present case.
The total pulling force can be calculated by 
\begin{equation}
\bm{F}_{\rm pull} = \sum_{i \in \cellJ} {(-{{\bm{D}_i}^{T}}\cdot \bm{\lambda })} = -\bm{\lambda} . 
\end{equation}

\begin{figure*}[t!]
\includegraphics[width=17.8cm]{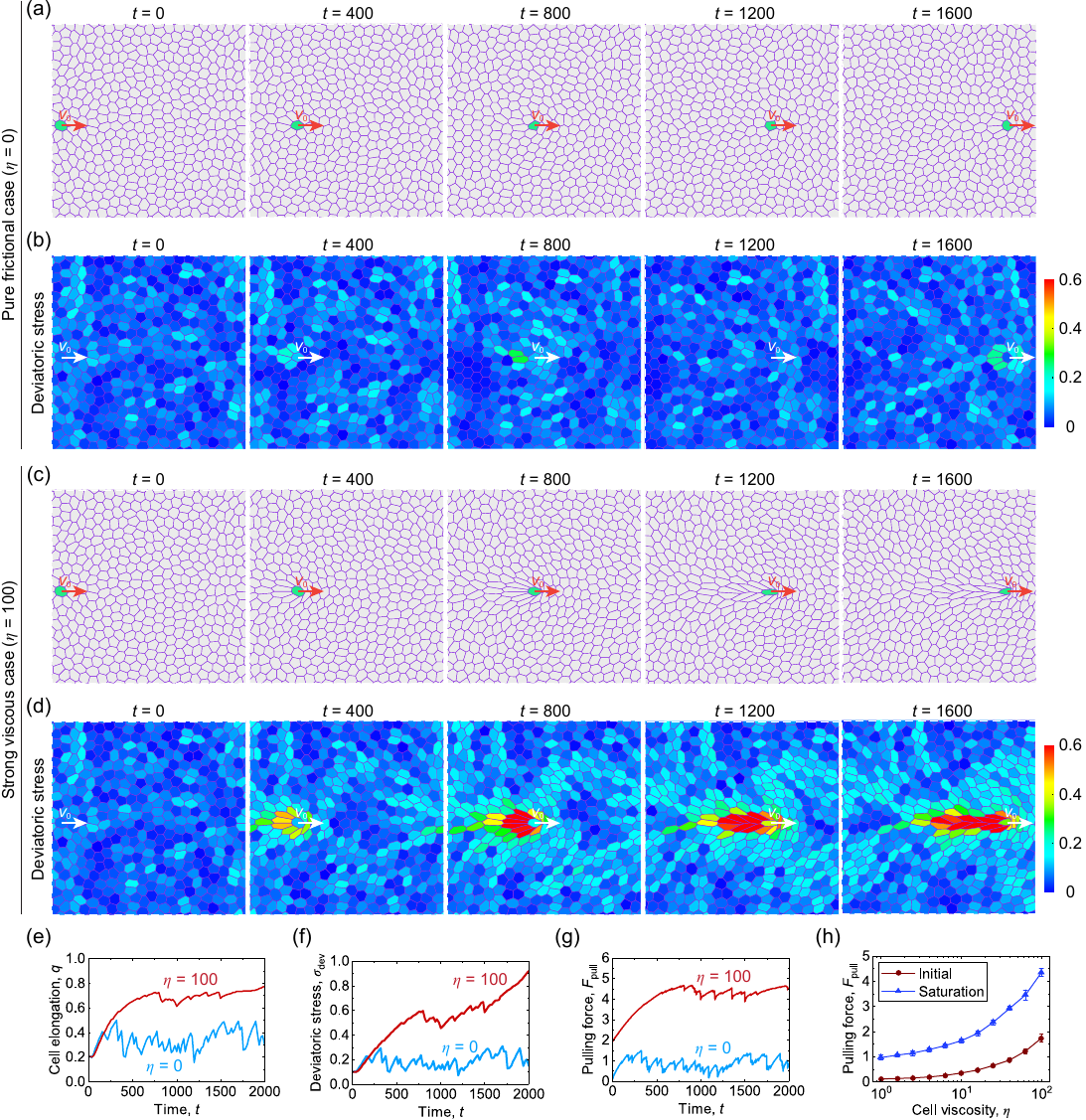}
\caption{\label{fig:pull_one_cell}
Numerical simulation of pulling a cell within a passive cell sheet with a constant pulling velocity $v_0$. Here, we do not consider cellular activities and assume $\eta_s = \eta_b = \eta$. 
Comparison of (a, b) the pure frictional case ($\eta = 0$) and (c, d) the strong viscous case ($\eta = 100$). 
(a, c) Evolution of cell morphologies. Here, we mark the cell under pulling in green. 
(b, d) Evolution of the deviatoric stress magnitude $\sigma_{\rm dev}$ within cells. 
The deviatoric stress magnitude is defined as $\sigma_{\rm dev} = | \bm{\sigma}_{\rm dev} |$, where $\bm{\sigma}_{\rm dev} = \bm{\sigma} - \sigma_{\rm iso}\bm{I}$ with $\sigma_{\rm iso} = {\rm tr}(\bm{\sigma}) / 2$ being the isotropic stress. 
(e) Comparison of the average cell elongation (of neighboring cells of the cell under pulling) for the pure frictional case and the strong viscous case. 
(f) Comparison of the average deviatoric stress magnitude (of neighboring cells of the cell under pulling) for the pure frictional case and the strong viscous case. 
(g) Comparison of the pulling force for the pure frictional case and the strong viscous case. 
(h) The pulling force $F_{\rm pull}$ as a function of the cell viscosity $\eta$. 
Here, we show the initial pulling force $F_{\rm pull}(t = 0)$ and the saturation pulling force $F_{\rm pull}(t = +\infty)$. 
Parameters: $T_0 = 0$, $\beta = 0$, $v_0 = 0.01$, and $\gamma = 1$. 
}
\end{figure*}

\subsubsection{Results}

We perform numerical simulations of pulling a cell within a passive cell sheet ($T_0 = 0$ and $\beta = 0$) with a constant drag velocity $v_0 = 0.01$. 
Here, we set $\gamma = 1$ and $\eta_s = \eta_b = \eta$. 
We compare two limiting cases: (1) pure frictional case ($\eta = 0$); (2) strong viscous case ($\eta = 100$). 

In Fig. \ref{fig:pull_one_cell}(a--d), we show representative cell morphologies and the corresponding deviatoric stress fields for the two cases. In the pure frictional case ($\eta = 0$), the cell deformation and the cell stress are relaxed (Fig. \ref{fig:pull_one_cell}(a, e, f)). 
In contrast, in the strong viscous case ($\eta = 100$), cells around the cell under pulling exhibit large deformations (Fig. \ref{fig:pull_one_cell}(c, e)), leading to large stresses (Fig. \ref{fig:pull_one_cell}(d, f)). 
To overcome such large cell deformations, a much larger pulling force $F_{\rm pull}$ is required for the strong viscous case compared to the pure frictional case (Fig. \ref{fig:pull_one_cell}(g)). 

Yet the scaling of $F_{\rm pull}$ is strongly sublinear in $\eta_{\mathrm{tissue}}^{\rm (LT)}$ (Fig. \ref{fig:pull_one_cell}(h)), in sharp contrast to the classical Stokes drag (as regularized by Oseen correction in 2D) for which the force remains proportional to the viscosity \cite{childress2009theoreticalfluid}. Understanding the origin of this sublinear scaling represents an interesting direction for future work.

\begin{figure}[t!]
\includegraphics[width=8.3cm]{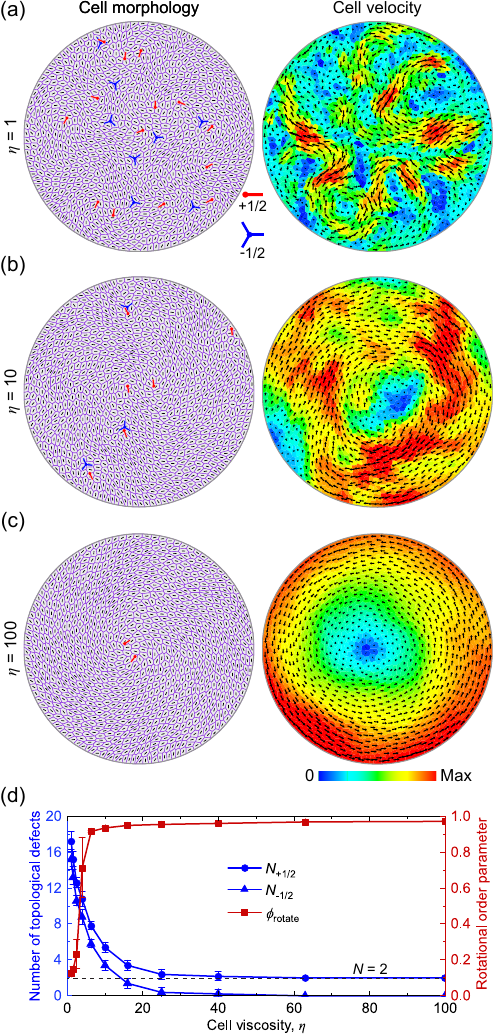}
\caption{\label{fig:circular_geometry}
Numerical simulation of viscous, polar, active tissue flows in a circular domain. 
(a-c) The cell morphology (\textit{left}) and the cell velocity field (\textit{right}) at various levels of cell viscosity $\eta_b = \eta_s = \eta$: (a) $\eta = 1$; (b) $\eta = 10$; (c) $\eta = 100$. 
The black lines represent cell orientation and the red (resp. blue) symbols indicate the locations and orientations of $+1/2$ (resp. $-1/2$) topological defects, extracted using the scheme proposed in Ref. \cite{Lin2023}.  
The arrows represent cell velocity vectors, and the color code refers to the cell velocity magnitude. 
(d) The number of topological defects and the rotational order parameter as functions of the cell viscosity $\eta$, averaged over $n = 5$ independent simulations. 
Parameters: $T_0 = 0.05$, $D_r = 0.05$, $\mu_{\rm LA} = 0.05$, and $\mu_{\rm CIL} = 1$.}
\end{figure}

\subsection{Polar active traction}
\label{sec:activity_polar}

\subsubsection{Method}

Here, we consider the case of polar active traction forces, setting the active stress to zero ($\beta=0$). 
The polar active traction force of a cell is usually modeled as 
\begin{equation}
\bm{F}_J^{(\rm active)} = T_0 \bm{p}_J , \label{eq:polar_active_traction_force}
\end{equation}
where $T_0$ denotes the typical magnitude of the traction force and $\bm{p}_J$ is the direction vector (referred to as the polarization vector) \cite{Lin2018,Li2014,Bi2016,Barton2017,Yin2024}, as shown in Fig. \ref{fig:Model}(e). 
The active force at each vertex, induced by the active traction forces of the surrounding cells, is then approximated by averaging the active traction forces around the vertex, i.e., $\bm{F}_i^{\rm (active)} = \langle \bm{F}_{J}^{\rm (active)} \rangle_{J \in C_i}$. 

We consider a model where the polarization vector $\bm{p}_J = (\cos(\theta_J),\sin(\theta_J))$ evolves under the combined influence of intercellular social interactions -- e.g., local alignment (LA) and contact inhibition of locomotion (CIL) -- and diffusive noise \cite{Bi2016,Barton2017,Lin2018}. 
Following Ref. \cite{Lin2018}, we express the dynamic equation of $\theta_J$ as 
\begin{align}
\frac{\mathrm{d}\theta_J}{\mathrm{d}t} = \ & \mu_{\rm LA} \frac{1}{n_J} \sum_{K \in \rm neighbor} \sin (\theta_K^{(\rm vel)} - \theta_J) \notag \\
& + \mu_{\rm CIL} \frac{1}{n_J} \sum_{K \in \rm neighbor} \sin (\alpha_{J,K} - \theta_J) \notag \\
& + \sqrt{2 D_r} \xi_J^{R}(t) , \label{eq:pJ_evolution}
\end{align}
where $\mu_{\rm LA}$ and $\mu_{\rm CIL}$ represent the intensities of LA and CIL, respectively; $D_r$ refers to a rotational diffusion coefficient, and $\xi_J^{R}(t)$ are independent unit-variance Gaussian white noises. 
In detail, the first term in Eq. \eqref{eq:pJ_evolution} accounts for the tendency of a cell to follow the motion direction of its neighbors (equivalent to a Vicsek-type, local alignment interaction), where $\theta_K^{\rm (vel)} = \arg (\bm{v}_K)$ is the immediate motion direction of cell $K$ and the summation $\sum_{K \in \rm neighbor}$ is over all contacting neighbor cells of the $J$-th cell.
The second term in Eq. \eqref{eq:pJ_evolution} describes the tendency of a cell to move away from its neighbors upon contact, i.e., a repulsion interaction, where $\alpha_{J,K} = \arg (\bm{r}_J - \bm{r}_K)$ refers to the argument of the direction pointing from cell $K$ to cell $J$. 
The third term in Eq. \eqref{eq:pJ_evolution} accounts for the random rotational diffusion of cell polarization vectors.

We consider a disk-like geometry \cite{Lin2018, Sonam2023, Li2014} with boundary vertices allowed to slide freely along the boundary. Thus, the boundary condition reads: $\bm{v}_i \cdot \bm{n}_i = 0$ for all boundary vertices, where $\bm{n}_i$ represents the unit vector normal to the boundary at the vertex $i$. Introducing the Lagrange multiplier $\bm{\lambda }$, the overall motion equation can be expressed as in Eq. (\ref{eq:MotionEquation_SlabShear}) with $\bm{v}_b=0$ here.

\subsubsection{Results}

We perform numerical simulations in a circular domain with a diameter much larger than the intrinsic swirl size of the cell sheet in the vanishing-viscosity limit \cite{Lin2018}. 
Thus, at low cellular viscosity $\eta = \eta_s = \eta_b$, we observe turbulent-like flows with many motile topological defects, which are singular points in the cell orientation pattern (Fig. \ref{fig:circular_geometry}(a)). 
Increasing cellular viscosity results in enhanced spatial correlation in cell orientation (marked by a decrease in the number of topological defects) and an enlarged swirl size (Fig. \ref{fig:circular_geometry}(b)). 
At larger cellular viscosity, we observe a global disk-like rotation mode, accompanied by two rotating $+1/2$ topological defects, located in the center region of the circular domain (Fig. \ref{fig:circular_geometry}(c)). 
The mode transition is further verified by quantifying the number of topological defects $N_{\pm 1/2}$ and the rotational order parameter $\phi_{\rm rotate} = \langle (1/N_c) \sum_J \hat{\bm{v}}_J \cdot \bm{e}_{\varphi,J} \rangle_t$ with $\hat{\bm{v}}_J = \bm{v}_J / |\bm{v}_J|$ and $\bm{e}_{\varphi,J}$ the local circumferential direction at cell $J$.

\begin{figure}[t!]
\includegraphics[width=8.6cm]{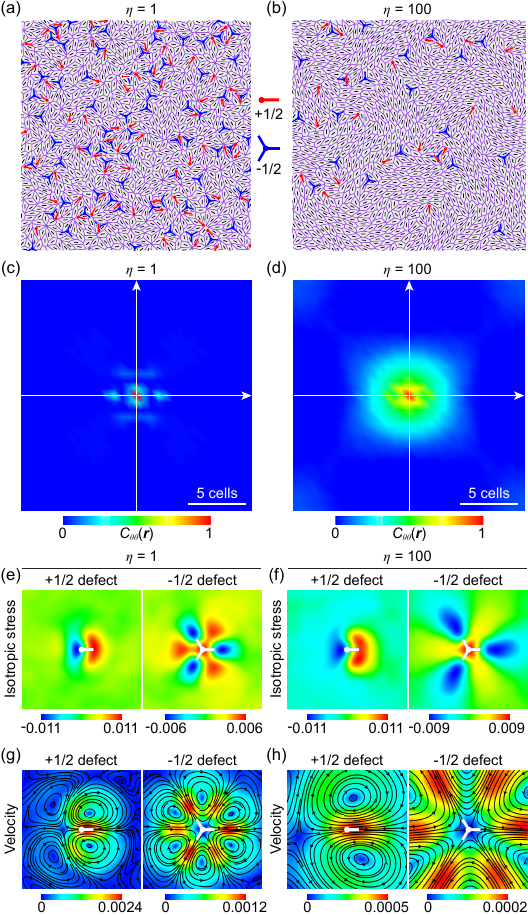}
\caption{\label{fig:TopologicalDefects_Viscosity}
Numerical simulation of a viscous, active nematic vertex model in a square domain with periodic boundary conditions. Here, we set $\eta_s = \eta_b = \eta$. 
(a, b) Typical cell morphologies at different cell viscosities: (a) $\eta = 1$; (b) $\eta = 100$. 
The black lines indicate cell orientations; the red (resp. blue) symbols represent $+1/2$ (resp. $-1/2$) topological defects, extracted using the scheme proposed in Ref. \cite{Lin2023}. 
(c, d) The spatial correlation map $C_{\theta\theta}(\bm{r})$ of cell shape orientation at different cell viscosities: (c) $\eta = 1$; (d) $\eta = 100$. Scale bar = 5 cell lengths. 
The spatial correlation function is defined as $C_{\theta\theta}(\bm{r}) = \langle\cos2(\theta_J - \theta_K)\rangle$, where $\langle\cdot\rangle$ averages over all cell pairs ($J$, $K$) satisfying $|(\bm{r}_J - \bm{r}_K) - \bm{r}| < \Delta r$ with $\Delta r = 0.5$. 
(e, f) The average isotropic stress and (g, h) the average flow field near topological defects at different cellular viscosities: (e, g) $\eta = 1$; (f, h) $\eta = 100$. 
In (e, f), the color code refers to the local stress fluctuation $\sigma_{\rm iso} - \langle \sigma_{\rm iso} \rangle$. 
In (g, h), the color code refers to the velocity magnitude, and the black lines with arrows represent flow directions. 
The number of defects for average: (e, g) $n_{\pm 1/2} = 135805$; (f, h) $n_{\pm 1/2} = 53307$. 
Domain size $L = 16$. 
Parameters: $T_0 = 0$ and $\beta = 0.4$.}
\end{figure}

\subsection{Nematic active stresses}
\label{sec:activity_nematic}

Here, we focus on a nematic active vertex model and turn off the active polar traction force ($T_0 = 0$). 

\subsubsection{Method}

Here, we consider the case of the apolar cellular active stress. 
In many cases, the apolar cellular active stress is anisotropic and depends on the cell shape \cite{Makhija2016,Singh2018,Uwamichi2024}. 
In the linear order, the cell-shape-dependent active stress can be modeled as
\begin{equation}
\bm{\sigma}_J^{(\rm active)} = - \beta \bm{Q}_J , 
\end{equation}
with $\beta$ being an activity parameter and $\bm{Q}_J$ a cell shape anisotropy tensor \cite{Lin2023,Sonam2023,Rozman2024,Rozman2025}, as shown in Fig. \ref{fig:Model}(f). 
The sign of $\beta$ quantifies whether a cell actively pulls or pushes its neighbors (Fig. \ref{fig:Model}(f)): when $\beta > 0$, a cell actively pushes its neighbors along its shape elongation direction; otherwise, a cell actively pulls its neighbors along its shape elongation direction. 

There are several different ways to define the cell shape anisotropy tensor, including the cell edge-based scheme, the cell vertex-based scheme, and the cell area-based scheme \cite{Lin2023}. 
These different ways lead to similar cell shape and pattern transitions \cite{Lin2023}. 
Here, using a cell edge-based scheme, we define the cell shape anisotropy tensor as in Ref. \cite{Lin2023}, 
\begin{equation}
\bm{Q}_J = \frac{1}{P_J} \sum_{k \in \cellJ} \ell_k \bm{t}_k \otimes \bm{t}_k - \frac{1}{2}\bm{I} , \label{eq:Q_definition}
\end{equation}
where $\ell_k = | \bm{r}_{k+1} - \bm{r}_k |$ and $\bm{t}_k = (\bm{r}_{k+1} - \bm{r}_k) / \ell_k$ are the length and direction of the $k$-th edge (pointing from vertex $k$ to vertex $k+1$) of the $J$-th cell. 
The cell shape anisotropy tensor $\bm{Q}_J$ is traceless and quantifies the elongation (corresponding to the positive eigenvalue of $\bm{Q}_J$) and orientation (corresponding to the eigenvector associated with the positive eigenvalue of $\bm{Q}_J$) of the $J$-th cell. 

The force at each vertex induced by the active stress $\bm{\sigma}_J^{(\rm active)}$ can be calculated using the Cauchy stress formula \cite{Tlili2019,Lin2022,Lin2023}, that is, by projecting the active stress of a cell onto its edges.

\subsubsection{Results}

Here, we consider a cell sheet consisting of $N_c = 1000$ cells in a square domain with periodic boundary conditions. 
We find that at low cell viscosity, strong active stresses lead to elongated cell shapes and multicellular rosettes, accompanied by a high density of topological defects, as shown in Fig. \ref{fig:TopologicalDefects_Viscosity}(a, c). 
This is consistent with our previous study of an active nematic vertex model at a vanishing viscosity limit \cite{Lin2023}. 
At high cell viscosity, we observe long-range spatial correlations in cell orientation and well-defined topological defects (Fig. \ref{fig:TopologicalDefects_Viscosity}(b, d)); these features are close to those observed in experiments \cite{Saw2017,Kawaguchi2017,Sonam2023,Balasubramaniam2021,Blanch-Mercader2018}. 

We further examine the isotropic stress field and the flow field near topological defects, averaged over more than 50000 topological defects (see Ref. \cite{Lin2023} for the average scheme), as shown in Fig. \ref{fig:TopologicalDefects_Viscosity}(e-h). 
These fields are consistent with active nematic gel theory and with reported experiments \cite{Saw2017,Balasubramaniam2021,Blanch-Mercader2018,Sonam2023}. 
As expected, a stronger cell viscosity results in stress and flow patterns of longer correlation length.

\subsubsection{Discussion}

Consistent with our previous findings \cite{Sonam2023}, we observe the formation of large-scale, multicellular topological defects. However, while these topological defects were previously driven by nematic alignment coupling between neighbors, they arise here as a result of cell viscosity. Furthermore -- although not detailed here -- our simulations yield spontaneous unidirectional flows in slab geometries under no-slip boundary conditions, in agreement with Ref. \cite{Rozman2025}.

\section{Conclusion}

Our work introduces a viscous extension of the vertex model in which dissipation arises from two microscopic sources: junctional viscosity along cell--cell interfaces and bulk viscous drag between vertices and cell centers. Both contributions are formulated in a rotationally invariant way, depending only on relative velocities projected along cell edges and vertex–cell segments, and assembled into a friction–viscosity coefficient matrix that governs the overdamped dynamics. A Lagrange multiplier framework is used to impose kinematic constraints, which regularizes the dynamics in the zero-friction limit and provides a unified way to implement fixed, sliding, and driven boundaries, as well as global momentum and angular-momentum conservation.

On this basis, a constrained slab-shear protocol is proposed to extract a coarse-grained tissue viscosity directly from vertex velocities and line tensions. 
An analytical calculation for a single hexagonal cell and numerical simulations of disordered packings show that the short-time tissue viscosity scales linearly with both junctional and bulk viscosities of cells, whereas sustained shear in frictionless systems allows for a long-time tissue viscosity to be defined in the presence of cell rearrangements. 
The same viscous vertex model is then applied to active polar and nematic tissues, where cell viscosity is shown to elongate and align cells, reduce the number density of topological defects, and reorganize active flows under confinement, driving a crossover from defect-rich active turbulence to coherent motion. 
Because the formulation remains well-posed at vanishing substrate friction and interfaces naturally with active nematic descriptions, it offers a useful bridge between cell-resolved simulations and continuum theories of free-standing tissues and organoids, and provides a framework for quantitatively interpreting future rheological measurements in living epithelia.

\section*{Acknowledgments}
S.Z.L. acknowledges support from the National Natural Science Foundation of China (Grant No. 12502367), Research Center for Magnetoelectric Physics of Guangdong Province (Grants 2024B0303390001), and Guangdong Provincial Key Laboratory of Magnetoelectric Physics and Devices (Grants 2022B1212010008).
J.-F.R. acknowledges support from France 2030, the French Government program managed by the French National Research Agency (ANR-16-CONV-0001) from Excellence Initiative of Aix-Marseille University - A*MIDEX. J.-F.R. and S.-Z.L. thank M. Merkel for interesting discussions. 
J.-F.R. and S.T. thank Nishchhal Verma for helpful initial simulations.

\bibliographystyle{apsrev4-2}
\bibliography{refs}

\clearpage

\appendix

\renewcommand{\thefigure}{A\arabic{figure}}

\setcounter{figure}{0}

\end{document}